\documentclass[sn-mathphys,Numbered]{sn-jnl}% Math and 

\usepackage{graphicx}%
\usepackage{dcolumn}
\usepackage{multirow}%
\usepackage{amsmath,amssymb,amsfonts}%
\usepackage{amsthm}%
\usepackage{mathrsfs}%
\usepackage[title]{appendix}%
\usepackage{xcolor}%
\usepackage{textcomp}%
\usepackage{manyfoot}%
\usepackage{booktabs}%
\usepackage{algorithm}%
\usepackage{algorithmicx}%
\usepackage{algpseudocode}%
\usepackage{listings}%
\usepackage{accents}
\usepackage{pdfpages}
\usepackage{booktabs}  % For professional looking tables
\usepackage{caption}   % For caption customization
\usepackage{array}     % For better column options
\usepackage[absolute,overlay]{textpos}
\usepackage{bm}
\usepackage{amsmath}
\usepackage{graphicx}
\usepackage{subcaption}
\usepackage{booktabs}
\usepackage{multirow}
\usepackage{makecell}
\usepackage{array}
\newcolumntype{P}[1]{>{\raggedright\arraybackslash}p{#1}}
\usepackage{enumitem}
\usepackage{booktabs} 
\usepackage{caption} 
\usepackage{float} 
\usepackage{mwe}
\theoremstyle{thmstyleone}%

\theoremstyle{thmstyletwo}%

\theoremstyle{thmstylethree}%

\begin{document}

%\title[Article Title]{A spatiotemporal deep neural network framework for predicting fracture energy and initiation and propagation of cracks in highly heterogeneous concrete microstructures}

\title[Article Title]{A spatiotemporal deep learning framework for prediction of crack dynamics in heterogeneous solids: efficient mapping of concrete microstructures to its fracture properties}

%\title[Article Title]{Enhancing reduced-order finite element output resolution through neural network integration: A comparative study of various approaches.}

\author*[1]{\fnm{Rasoul} \sur{Najafi Koopas}}\email{najafikr@hsu-hh.de}

\author[2]{\fnm{Shahed} \sur{Rezaei}}\email{s.rezaei@access-technology.de}

\author[1]{\fnm{Natalie} \sur{Rauter}}\email{natalie.rauter@hsu-hh.de}

\author[1]{\fnm{Richard} \sur{Ostwald}}\email{ostwald@hsu-hh.de}
\author[1]{\fnm{Rolf} \sur{Lammering}}\email{rolf.lammering@hsu-hh.de}

\affil*[1]{\orgdiv{Chair of Solid Mechanics}, \orgname{Helmut-Schmidt University/University of the Federal Armed Forces}, \orgaddress{\street{Holstenhofweg  85}, \city{Hamburg}, \postcode{22043}, \country{Germany}}}

\affil[2]{Access e.V., Aachen, Germany}

\abstract{ 
A spatiotemporal deep learning framework is proposed that is capable of two-dimensional full-field prediction of fracture in concrete mesostructures. This framework not only predicts fractures but also captures the entire history of the fracture process, from the crack initiation in the interfacial transition zone (ITZ) to the subsequent propagation of the cracks in the mortar matrix. Additionally, a convolutional neural network (CNN ) is developed which is capable of predicting the averaged stress-strain curve of the mesostructures. The UNet modeling framework, which comprises an encoder-decoder section with skip connections, is used as the deep learning surrogate model. Training and test data are generated from high-fidelity fracture simulations of randomly generated concrete mesostructures. These mesostructures include geometric variabilities such as different aggregate particle geometrical features, spatial distribution, and the total volume fraction of aggregates. The fracture simulations are carried out in Abaqus/CAE, utilizing the cohesive phase-field fracture modeling technique as the fracture modeling approach. In this work, to reduce the number of training datasets, the spatial distribution of three sets of material properties for three-phase concrete mesostructures, along with the spatial phase-field damage index, are fed to the UNet to predict the corresponding stress and spatial damage index at the subsequent step. It is shown that after the training process using this methodology, the UNet model is capable of accurately predicting damage on the unseen test dataset by using just 470 datasets. Moreover, another novel aspect of this work is the conversion of irregular finite element data into regular grids using a developed pipeline. This approach allows for the implementation of less complex UNet architecture and facilitates the integration of phase-field fracture equations into surrogate models for future developments.
}


\maketitle
%%%%%%%%%%%%%%%%%%
%%%%%%%%%%%%%%%%%%
%%%%%%%%%%%%%%%%%%
%%%%%%%%%%%%%%%%%%
\newpage
\section{Introduction}\label{sec1}
Numerical multiscale fracture simulations pose a significant challenge even with modern high-performance computers, as a large system of equations has to be solved for each time step. Achieving convergence at each step is extremely time-consuming and computationally expensive. Additionally, concrete microstructures include a phase known as the interfacial transition zone (ITZ), which has poor mechanical properties. During loading scenarios involving fractures, damage typically initiates in the ITZ and then propagates into the mortar matrix, further complicating simulations and increasing computational demands. One possible solution to reduce computational costs is the implementation of surrogate models, which facilitate the application of concurrent multiscale frameworks by significantly reducing the simulation time at each subscale \cite{gupta2023accelerated, li2024machine}.

Many of the surrogate models developed in the literature focus on predicting the homogenized quantities of loaded representative volume elements (RVEs) \cite{li2019predicting, yang2018deep, yang2020prediction, rao2020three, kim2023surrogate, yang2023predicting, mendikute2023predicting}. However, a full-field spatiotemporal prediction of certain local quantities, such as damage or stress fields, is crucial in multiscale modelings. For instance, in concrete materials, a full-field prediction of the damage field at the meso- or microscale enables the identification of weak spots in large concrete structures, such as bridges, with high accuracy. This, in turn, makes structural health monitoring of such structures more effective due to precise sensor positioning. Additionally, a full-field spatiotemporal prediction facilitates the study of the effects of different microstructural and material property features on crack patterns. This enables microstructure optimization, thereby achieving a composite material with superior properties \cite{liu2023multiscale}.

Implementing machine learning and deep learning approaches to generate high-fidelity predictions is advancing rapidly. Below, we briefly highlight some comprehensive works among many contributions that demonstrate the potential of machine learning approaches in the full-field prediction of the behavior of composite materials. \citet{wang2024efficient} introduced a surrogate model for predicting full-field damage in laminated composite structures under low-velocity impacts, using advanced Vector Quantised-Variational AutoEncoder. \citet{chen2023full} developed a surrogate framework using two sequential CNNs for predicting stresses and cracks in composite material microstructures, where the inclusion of a self-attention layer enhances the model's ability to capture stress and fracture patterns. \citet{CHANG2022108624} proposed a CNN-based method for analyzing fractures in printed concrete structures. This work predicts stress-crack width curves by integrating principal component analysis with a U-Net framework that predicts crack patterns. Additionally, transfer learning is used for model generalization to varied microstructures. \citet{DING2022116248} developed two back-propagation deep neural network models where the first one is a regression model for predicting the macroscopic transverse mechanical properties of fiber-reinforced polymers based on Discrete Element Method simulations of RVEs, and the second one is a classification model for predicting microscopic crack patterns. \citet{yan2023spatiotemporal} developed an end-to-end spatiotemporal prediction model with Encoder-Translator-Decoder architecture to predict delamination growth under compressive loading in composite samples a single hole. Additionally, a transfer framework with a coupling coding block could predict multi-hole delamination growth, enhancing training speed and prediction accuracy. \citet{mohammadzadeh2022predicting} extended the Mechanical MNIST dataset \cite{lejeune2020mechanical} to include finite element simulation results of quasi-static brittle fracture in heterogeneous materials using the phase-field fracture approach. In this study, various deep neural network architectures were tested, and the MultiRes-WNet architecture showed strong baseline performance on the Mechanical MNIST Crack Path dataset. \citet{sepasdar2021data} developed an image-based deep learning framework, enhanced by a physics-informed loss function, to predict damage and failure in composites. The framework consists of two sequentially trained fully-convolutional networks: Generator 1, which predicts post-failure stress distribution, and Generator 2, which predicts the failure pattern. Moreover, CNNs have been used to predict full-field prediction of local variables of interest in various materials \cite{yan2023multi, ding2023effects, croom2022deep, bhaduri2022stress, yang2021deep, jiang2021stressgan, nie2020stress}. For example, \citet{kim2023data} introduced a deep neural network approach to predict stress–strain curves for unidirectional composites. By generating RVEs and using finite element simulations, the model combined with principal component analysis accurately predicted stress-strain curve with about 2\% error in toughness. \citet{shokrollahi2023deep} implemented a deep learning framework for predicting stress fields in composite materials, regardless of geometric complexity or boundary conditions. A UNet architecture is employed to correlate spatial fiber organization with resultant von Mises stress fields.

The aforementioned contributions clearly demonstrate the importance and effectiveness of new deep-learning-based surrogate models for fracture analysis, particularly in inhomogeneous and composite materials. However, significant shortcomings still need to be addressed. First, these studies are applied to relatively simple microstructures, neglecting the influence of the interface zone entirely. Yet, we know that crack patterns and the overall fracture toughness of a material are strongly affected by such interfacial zones, and the geometrically localized nature of cracks should be studied properly \cite{koopas2023comparative}. Additionally, most of the reviewed works either do not consider the history of damage propagation or, if they do, they implement LSTM in their surrogate framework. As mentioned in \cite{chen2023full}, the use of the long short-term memory (LSTM) networks may not be ideal for predicting crack initiation and propagation because the spatiotemporal characteristics of the microstructure learned from memory remain consistent over the entire time domain.

In this work, we developed a spatiotemporal UNet-based surrogate model capable of predicting the average stress-strain curve and final crack pattern of concrete mesostructures. The novelty of the developed framework lies in the following aspects: (i) the model is trained on a dataset of concrete mesostructures containing ITZ with various geometrical attributes of aggregate particles, their volume fraction, and spatial distributions; (ii) the framework considers the history of fracture, fully capturing the initiation of cracks in the ITZ and their propagation in the mortar matrix; (iii) it predicts the average stress-strain curve and fracture energy, enabling future mesostructures optimization in terms of fracture energy; (iv) instead of implementing image-to-image mapping, the input to the UNet contains three different sets of spatial distributions of material properties (modulus of elasticity, ultimate tensile strength, and fracture energy) along with the spatial cohesive phase-field damage index, thus requiring very few training data for the deep learning framework; (v) the dataset is generated by implementing cohesive phase-field fracture FE simulations using Abaqus/CAE on complex concrete mesostructures, where a free meshing technique is implemented for discretization. Consequently, a specially designed pipeline interpolates the FE data to a regular grid with high accuracy, eliminating the need for complex surrogate frameworks like graph neural nets and making it possible to implement neural operator learning on the current model to increase accuracy for future development.

\section{Problem description} 
\begin{figure}[t]
\center
    \includegraphics[width=0.7\textwidth]{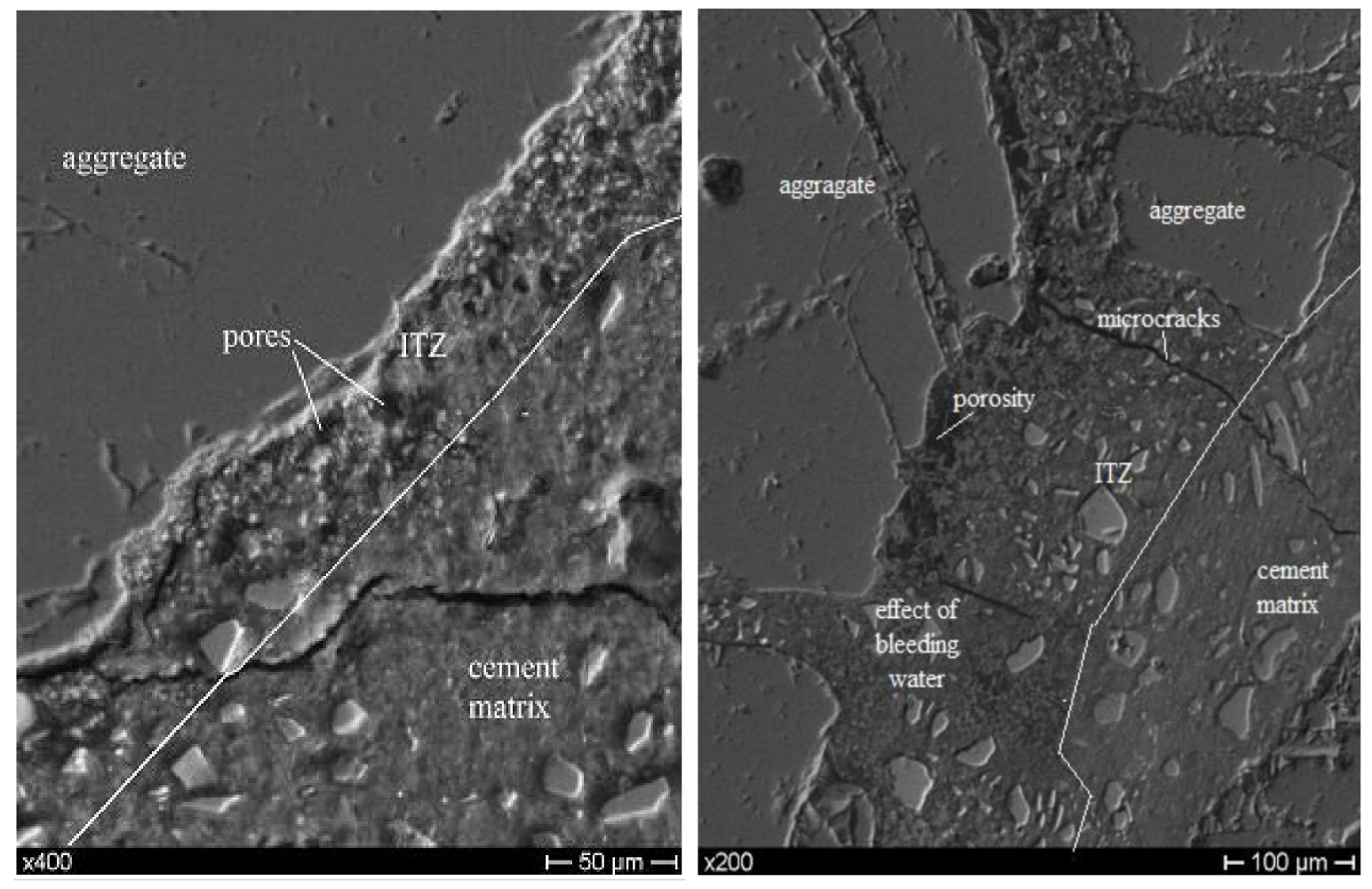}
    \caption{SEM imaging of interface zone in concrete materials \cite{bonifazi2015itz}. The interface zone in concrete materials has a thickness between 20-100 µm, which, compared to other phases, is quite small and has relatively weak material properties.}
    \label{ITZ_elec}
\end{figure}

\begin{figure}[t]
\center
    \includegraphics[width=0.7\textwidth]{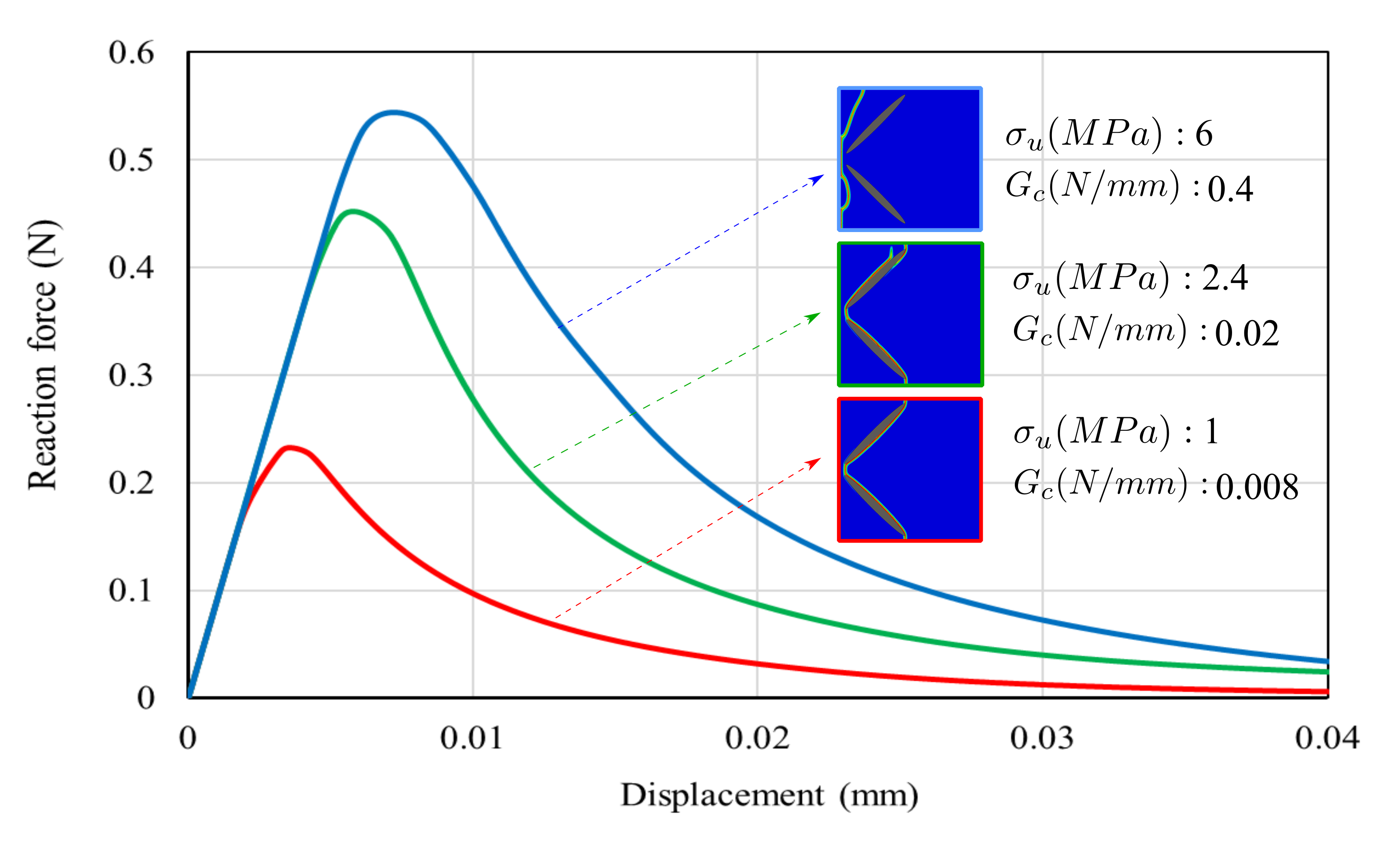}
    \caption{Effect of ITZ fracture properties on the fracture toughness of the concrete microstructure \cite{koopas2023comparative}.}
    \label{ITZ_Motiv}
\end{figure}

The interfacial transition zone significantly influences the macroscopic properties of concrete materials. As depicted in Fig.~\ref{ITZ_elec}, this comparably small zone is characterized by high porosity and consequently represents weak material properties. Under applied load conditions, microcracks first initiate within this zone before propagating into the mortar matrix. Therefore, a reliable and accurate modeling of concrete fracture requires considering this critical aspect. Fig.~\ref{ITZ_Motiv} illustrated the influence of ITZ fracture properties on the overall fracture toughness of the concrete microstructure. As mentioned, the innovative aspect of this work is the development of a surrogate model that is capable of capturing crack initiation within the ITZ and its subsequent propagation into the mortar matrix. Additionally, this model can also simulate other phenomena such as crack bridging. The training dataset for the surrogate model, representing these properties, is obtained through finite element fracture simulations of concrete mesostructures. The cohesive phase-field fracture model introduced by \citet{wu2017unified} is employed as the fracture modeling technique, which is suitable for small-scale simulations \cite{koopas2023comparative, REZAEI2022108177}. In the following, the theoretical aspects of this fracture model are briefly explained.

\subsection{Phase-field fracture model for brittle fracture}
\begin{figure}[t]
\center
    \includegraphics[width=0.8\textwidth]{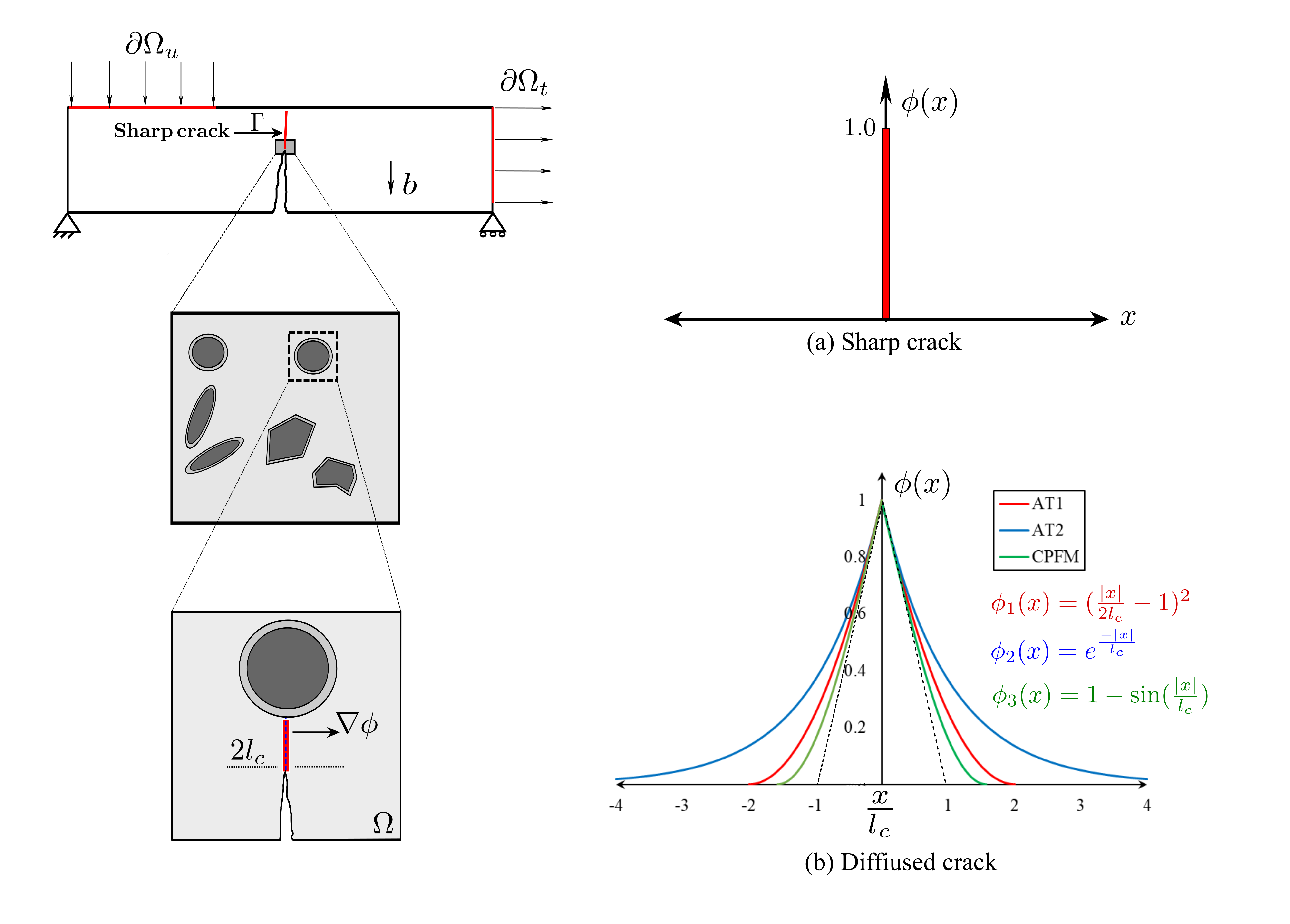}
    \caption{ The solid domain \(\Omega\) exhibits a diffuse crack, where the width of the diffuse band is quantified by the length scale parameter \(l_{c}\).}
    \label{domain}
\end{figure}

The phase-field fracture methodology approximates sharp cracks with diffused bands. In Fig.~\ref{domain}, domain \( \Omega \) features a diffused crack width defined by the length scale parameter \( l_{c} \). As \( l_{c} \) decreases, the model approaches a sharp crack. Additionally within Fig.~\ref{domain}, Neumann and Dirichlet boundary conditions are specified as \( \Omega_{t} \) and \( \Omega_{u} \) respectively.

To illustrate the regularization of a sharp crack topology, Fig.~\ref{domain} depicts a one-dimensional diffuse crack at \( x = 0 \) using
\begin{equation}
  \begin{cases}
    \phi_{1}(x) = \left(\frac{|x|}{2\,l_{c}}-1 \right)^2,\\[8pt]
    \phi_{2}(x) = e^{\frac{-|x|}{l_{c}}},\\[8pt]
    \phi_{3}(x) = 1-\sin\left(\frac{|x|}{l_{c}}\right)
  \end{cases}
  \label{strong1}
\end{equation}
where this model constrains \( \phi(x) \) such that
\begin{equation}  
    \phi(x=0)=1, \quad \phi(x) \to 0 \text{ as } x \to \pm\infty.
\end{equation}
Within the context of the phase-field fracture method, \(\phi \in [0, 1]\) denotes the degree of damage, where \(\phi = 0\) represents an undamaged state and \(\phi = 1\) represents complete damage. As shown in Eq.~\ref{irr_damage}, the material time derivative of the phase field damage parameter must remain non-negative to ensure the irreversibility of the damage, i.e.
\begin{equation}  
   \dot{\phi}(\mathbf{x} ,t) \geq 0.
   \label{irr_damage}
\end{equation}

The internal potential energy of the solid, following \cite{bourdin2008variational}, takes the form
\begin{equation}  
   \psi(\bm{\varepsilon}, \phi, \nabla_{\! \underaccent{\tilde}{\bold{x}}} \phi) = \psi_{e}(\bm{\varepsilon}, \phi) + \psi_{c}(\phi, \nabla_{\!\underaccent{\tilde}{\bold{x}}} \phi),
\end{equation}
where \(\nabla_{\!\underaccent{\tilde}{\bold{x}}}\) is the spatial gradient operator, \( \bm{\varepsilon} \) represents the linear strain tensor, \( \psi_{e} \) the stored strain energy, and \( \psi_{c} \) the fracture surface energy. The stored strain energy depends on the damage state in terms of
\begin{equation}  
    \psi_{e}(\bm{\varepsilon},\phi) = \omega(\phi)\,\psi_{0}(\bm{\varepsilon}),
    \label{elastic_energy}
\end{equation}
with the damage degradation function \( \omega(\phi) \) modulating material stiffness with the constraints
\begin{equation}  
    \omega(\phi=0)=1, \quad \omega(\phi=1)=0, \quad \omega'(\phi=1)=0.
    \label{constrains}
\end{equation}

Within Eq.~\ref{elastic_energy}, $\psi_{0}$ represents the strain energy density function of an undamaged solid. For an isotropic elastic body under material and a geometrically linear setting, $\psi_{0}$ is given by
\begin{equation}
    \psi_{0}(\bm{\varepsilon}) = \frac{1}{2}\,\bm{\varepsilon}:\mathbb{E}_{0}:\bm{\varepsilon} = \frac{1}{2}\,\lambda_{0}\,\mathrm{tr}^2\,(\bm{\varepsilon})+\mu_{0}\,\bm{\varepsilon}:\bm{\varepsilon}=\frac{1}{2}\,\bm{\overline{\sigma}}:\mathbb{C}_{0}:\bm{\overline{\sigma}}=\psi_{0}(\bm{\overline{\sigma}}),
    \label{HFE}
\end{equation}
where $\mathbb{E}_{0}=2\,\mu_{0}\,\mathbb{I}_{0}+\lambda_{0}\,\bm{I}\otimes\bm{I}$ is the fourth-order elastic stiffness tensor, $\lambda_{0}$ and $\mu_{0}$ are the Lamé constants, and $\mathbb{I}_{0}$ and $\bm{I}$ are the fourth-order and second-order identity tensors, respectively. Here, $\mathbb{C}_{0}=\mathbb{E}_{0}^{-1}$ is the compliance tensor, and $\bm{\overline{\sigma}}=\mathbb{E}_{0}:\bm{\varepsilon}$ is the effective stress tensor.

The fracture surface energy $\psi_{c}$ is expressed as a volume integral of the form \cite{verhoosel2013phase}:
\begin{equation}
    \psi_{c}(\phi, \nabla_{\!\underaccent{\tilde}{\bold{x}}} \phi) = \int_\Gamma G_{c} \,\mathrm{d}\Gamma  \approx \int_\Omega G_{c}\gamma(\phi, \nabla_{\!\underaccent{\tilde}{\bold{x}}} \phi) \,\mathrm{dA},
\end{equation}
where $G_{c}$ is the critical energy release rate, characterizing the material's fracture toughness. The function $\gamma(\phi, \nabla_{\!\underaccent{\tilde}{\bold{x}}} \phi)$ represents the fracture surface density, computed as
\begin{equation}
    \gamma(\phi, \nabla_{\!\underaccent{\tilde}{\bold{x}}} \phi) =  \frac{1}{c_{0}}\left[\frac{1}{l_{c}}\,\alpha (\phi) + l_{c}\nabla_{\!\underaccent{\tilde}{\bold{x}}} \phi \cdot \nabla_{\!\underaccent{\tilde}{\bold{x}}} \phi\right],
    \label{surf_dens_func}
\end{equation}
where $\alpha (\phi)$ is the crack geometric function, primarily employed to simulate the uniform evolution of the phase-field fracture parameter $\phi$. Additionally, the parameter $c_{0}$ is introduced to scale the regularized $\psi_{c}(\phi, \nabla_{\!\underaccent{\tilde}{\bold{x}}} \phi)$. 

As a result, the internal potential energy $\psi(\bm{\varepsilon}, \phi, \nabla_{\!\underaccent{\tilde}{\bold{x}}} \phi)$ takes the form
 
\begin{equation}  
   \psi(\bm{\varepsilon}, \phi, \nabla_{\!\underaccent{\tilde}{\bold{x}}} \phi) =  \int_{\Omega}^{} \omega(\phi) \, \psi_{0}(\bm{\varepsilon}) \,\mathrm{dV} + \int_{\Omega}^{} \frac{G_{c}}{c_{0}} \left [\frac{1}{l_{c}}\,\alpha (\phi) + l_{c}\nabla_{\!\underaccent{\tilde}{\bold{x}}} \phi\cdot\nabla_{\!\underaccent{\tilde}{\bold{x}}} \phi\right]  \,\mathrm{dV}.
\end{equation}
When evaluating the influence of externally applied loads on the solid domain denoted as $\Omega$, the computation of the external potential energy is undertaken as
\begin{equation}
    \varPi_{\mathrm{ext}} = \int_\Omega \bm{b^*} \cdot \bm{u} \,\mathrm{dV}  + \int_{\partial\Omega} \bm{t^*}\cdot \bm{u} \,\mathrm{dA},
\end{equation}
where $\bm{b^*}$ represents the volume forces, while $\bm{t^*}$ signifies the boundary forces. Consequently, the total potential energy of the solid domain is expressed as

\begin{equation}
    \varPi_{\mathrm{total}} = \int_{\Omega} \omega(\phi)\,\psi_{0}(\bm{\varepsilon}) \,\mathrm{dV} + \int_{\Omega} \frac{G_{c}}{c_{0}}\left[\frac{1}{l_{c}}\,\alpha (\phi) + l_{c}\,\nabla_{\!\underaccent{\tilde}{\bold{x}}} \phi \cdot \nabla_{\!\underaccent{\tilde}{\bold{x}}} \phi \right] \,\mathrm{dV} - \int_\Omega \bm{b^*} \cdot \bm{u} \,\mathrm{dV} - \int_\Omega \bm{t^*} \cdot \bm{u} \,\mathrm{dA}.
    \label{energy_functional}
\end{equation}

Eq.~\ref{energy_functional} represents the quadratic and convex total energy functional \cite{wu2017unified}. Minimizing $ \varPi_{\text{total}}$ yields the displacement and phase-field fracture parameters $(\bm{u}, \phi)$ via
\begin{equation}
\label{energy_min}
    (\bm{u}, \phi) = \mathrm{arg}\{\mathrm{min}\, \varPi_{\mathrm{total}}(\bm{u}, \phi)\} \quad \text{subject to}\quad \dot\phi \geq 0 \quad \text{and} \quad \phi \in [0,1].
\end{equation}

By applying the first variation of $\varPi_{\text{total}}$ and the divergence theorem, the strong form of the governing equation for the displacement follows as
\begin{equation}
  \begin{cases}
    \nabla \cdot \bm{\sigma} + \bm{b^*} = 0, \\
    \bm{u} = \bm{u}_{0} \quad \text{on } \partial\Omega_{u},\\
    \bm{\sigma} \cdot \bm{n} = \bm{t^*} \quad \text{on } \partial\Omega_{t},
  \end{cases}
  \label{strong2a}
\end{equation}
while the strong form of the equation governing the evolution of the  phase-field variable emerges as
\begin{equation}
  \begin{cases}
    Y - G_{c}\left[\frac{\partial \gamma(\phi, \nabla_{\!\underaccent{\tilde}{\bold{x}}} \phi)}{\partial \phi} - \nabla_{\!\underaccent{\tilde}{\bold{x}}} \cdot \left(\frac{\partial \gamma(\phi, \nabla_{\!\underaccent{\tilde}{\bold{x}}} \phi)}{\partial \nabla_{\!\underaccent{\tilde}{\bold{x}}} \phi}\right)\right] = 0 & \text{if } \dot \phi > 0, \\[5ex]
    Y - G_{c}\left[\frac{\partial \gamma(\phi, \nabla_{\!\underaccent{\tilde}{\bold{x}}} \phi)}{\partial \phi} - \nabla_{\!\underaccent{\tilde}{\bold{x}}} \cdot \left(\frac{\partial \gamma(\phi, \nabla_{\!\underaccent{\tilde}{\bold{x}}} \phi)}{\partial \nabla_{\!\underaccent{\tilde}{\bold{x}}} \phi}\right)\right] < 0 & \text{if } \dot \phi = 0,
  \end{cases}
  \label{strong2}
\end{equation}
where $Y$ is the energetic crack driving force, defined as
\begin{equation}
    Y = -\,\frac{\partial \psi}{\partial \phi} = -\,\omega'(\phi) \,\frac{\partial \psi}{\partial \omega(\phi)}.
\end{equation}

Since in this study the crack initiation and propagation in the concrete mesostructure are studied, the cohesive phase field fracture (CPFM) model introduced by \citet{wu2017unified} is used for FE simulations. A key feature of this modeling technique is its insensitivity to the length scale parameter $l_{c}$, making it suitable for small-scale simulations. The damage degradation function in the CPFM framework is given by \citet{conti2016phase}
\begin{equation}
    \omega(\phi) = \frac{(1-\phi)^2}{(1-\phi)^2 + a_{1}\,\phi + a_{1}\,a_{2}\,\phi^2 + a_{1}\,a_{2}\,a_{3}\,\phi^3},  
\end{equation}
where the constant variables $a_{1}, a_{2}$ and $ a_{3}$ are material dependent variables.
%############################################################################################
%############################################################################################
\section{Generation of training data}
To generate the data required for training a deep neural network to predict the initiation and propagation of cracks at the mesoscale in concrete, FE simulations are carried out using Abaqus/CAE. The process of generating appropriate data for neural network training can be categorized into three steps. In the first step, a pipeline, i.e. a sequential computational procedure, is developed to generate the required number of input data sets for mesoscale FE simulations of concrete materials. In the second step, FE simulations are conducted on the generated input data. In the final step, the FE simulation results are post-processed to make them suitable for the training phase of the neural network. This step is crucial as it enables the implementation of a spatiotemporal neural network framework that captures the initiation and propagation of cracks in concrete microstructures while managing the required memory and available GPU resources. Using the raw FE simulation results directly would be memory-intensive and potentially impractical, requiring a much more complex neural network architecture. The following section describes each step in detail.
\subsection{Generation of simulation input data}
\begin{figure}[H]
\center
    \includegraphics[width=\textwidth]{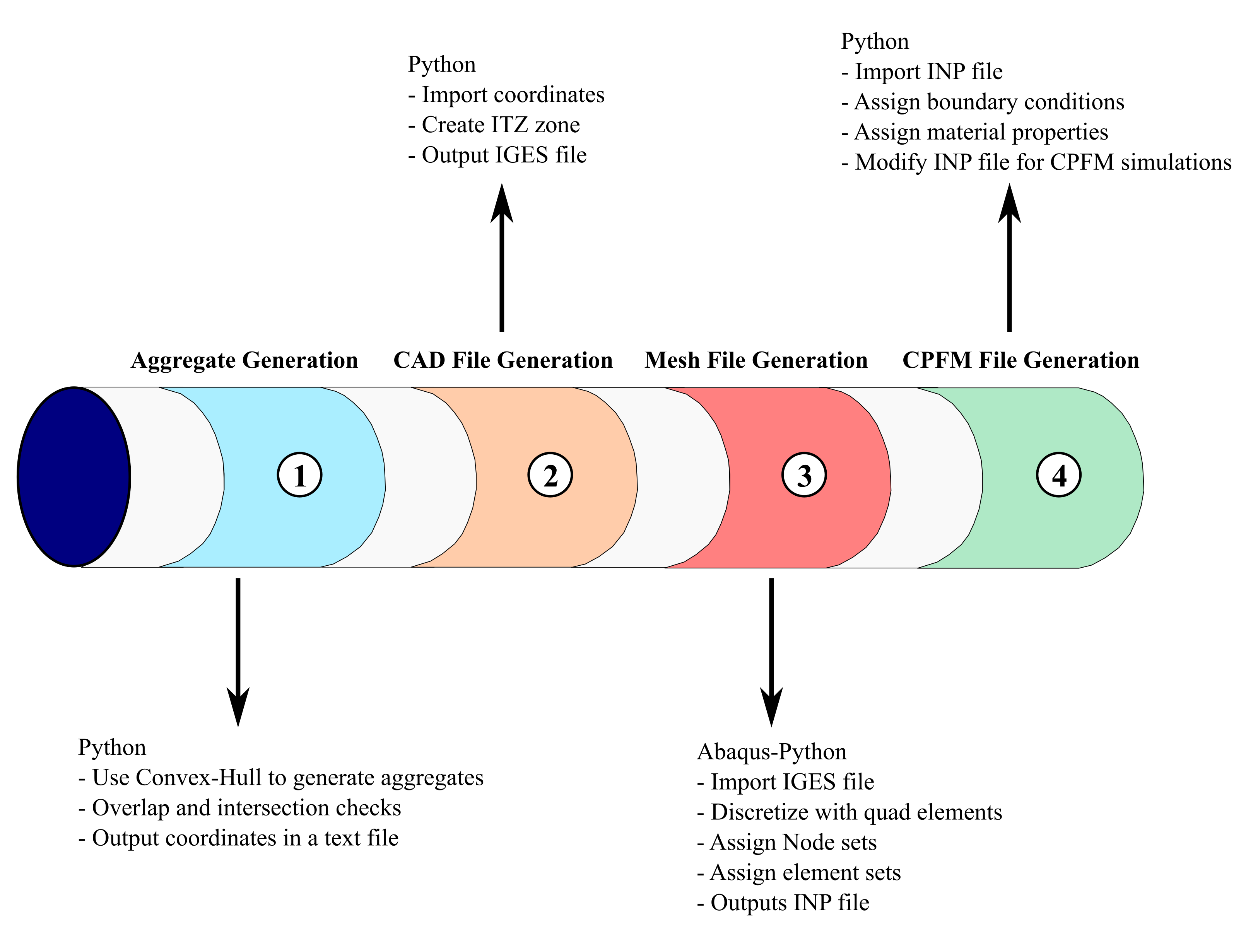}
    \caption{Fully-automated computational pipeline for generating input data for cohesive phase-field fracture simulations at the mesoscale for concrete materials.}
    \label{pipeline}
\end{figure}

To generate the input data for FE simulations, we developed a fully automated pipeline using Python. The user simply defines the desired number of simulations. Once specified, the pipeline saves the generated input data in distinct folders, where the name of each folder corresponds to its unique ID number within the pipeline. Fig.~\ref{pipeline} schematically represents the computational pipeline, its subcomponents, and the operations carried out in each subcomponent. 
\subsubsection{Aggregate particles generation}
\begin{figure}[H]
    \centering
    
    \captionsetup[subfigure]{justification=centering} % Center captions within subfigures

    \begin{subfigure}[b]{0.45\textwidth}
        \centering
        \includegraphics[width=\textwidth]{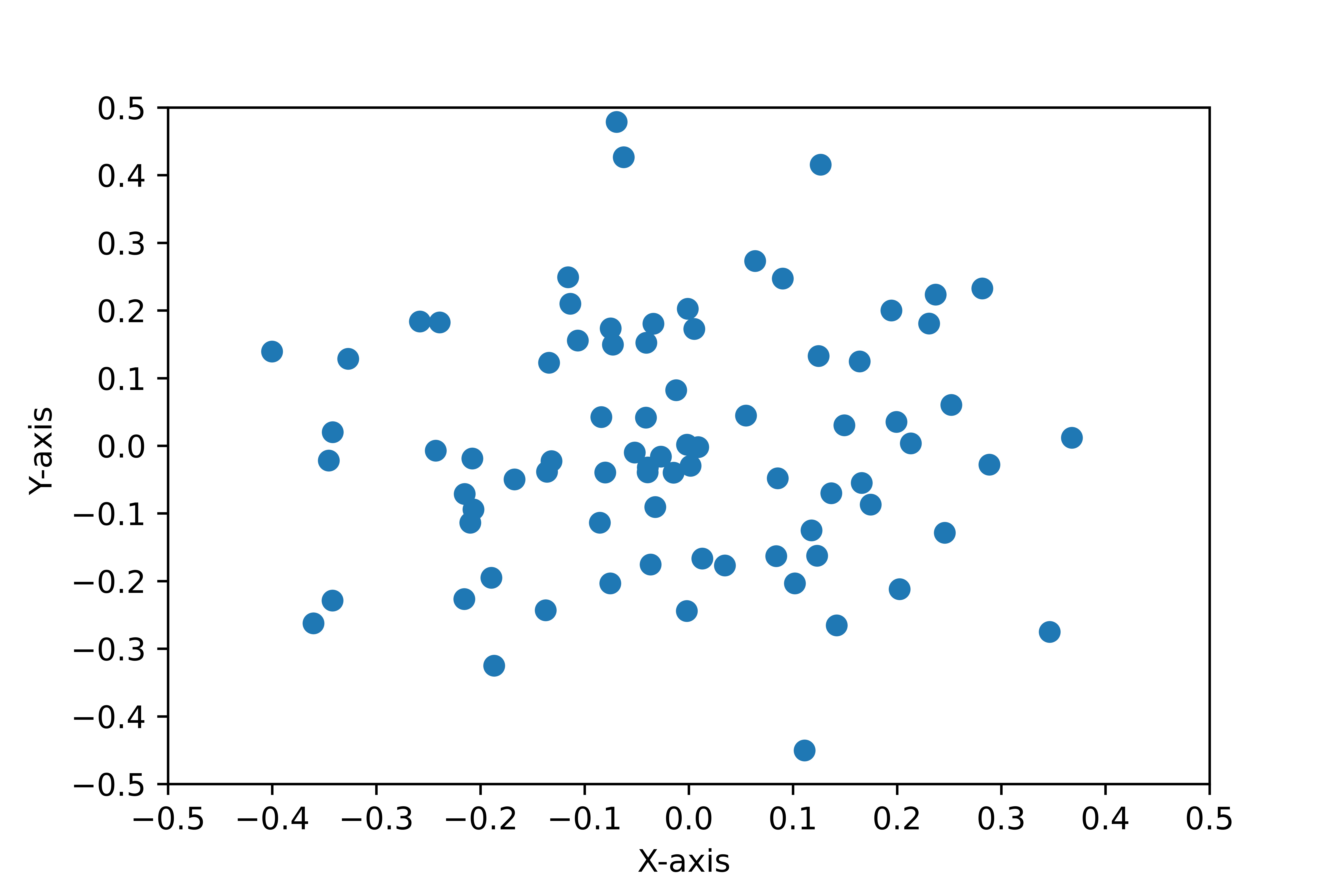}
        \caption{Random points}
        \label{fig:f1}
    \end{subfigure}
    \hfill
    \begin{subfigure}[b]{0.45\textwidth}
        \centering
        \includegraphics[width=\textwidth]{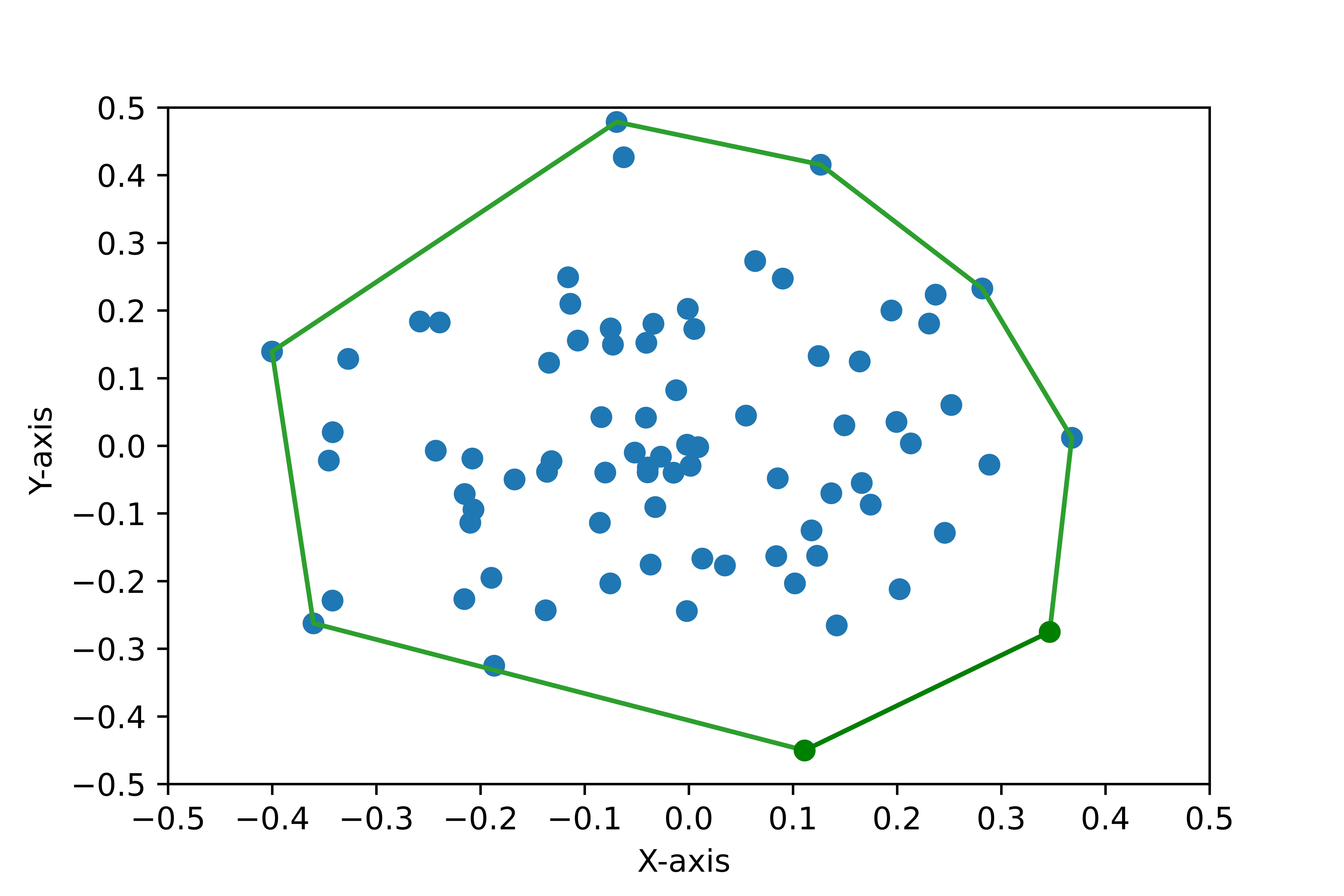}
        \caption{Convex hull of random points}
        \label{convexHull_randPoints}
    \end{subfigure}
    \caption{The generation of an aggregate particle is achieved by computing the convex Hull of a set of random points.}
    \label{fig:mainfig}
\end{figure}

As shown in Fig.~\ref{pipeline}, the first step in the pipeline designed for finite element simulation of mesoscale fracture in concrete is the generation of the concrete mesostructure, focusing on the generation of constructive points of aggregate particles. This phase is crucial as it forms the basis for accurately modeling the heterogeneous nature of concrete, which significantly affects its mechanical properties. The process begins with the creation of random points within the concrete mesoscale main frame that serve as centers for the aggregates. Using these centers, the code generates ellipses of various sizes and arbitrary angles and then selects a subset of random points from these ellipses to form the vertices of the aggregate polygons. It is therefore also possible to generate other types of aggregates such as elliptical or circular aggregates. The random ellipse sizes and angles are extracted from a user-defined distribution. The most important step here is the use of the Convex Hull algorithm to form a realistic polygon shape from these vertices. Fig.~\ref{convexHull_randPoints} illustrates the Convex Hull of a set of points randomly distributed in the plane \emph{XY}. Once the convex Hull is created, the code performs a series of intersection checks to ensure that the newly placed aggregate does not overlap with existing aggregates or extend beyond the main frame of the specimen. The intersection checks are carried out using the Python's \texttt{Shapely} library. The intersection checks to ensure that aggregates are within the specimen boundaries, do not overlap with each other, and maintain a small gap to create a mortar layer. For a thorough understanding of the process, interested readers are referred to \cite{najafi2023two}. Additionally, a flowchart outlining the process for generating the concrete microstructures is available Appendix A (see Fig.~\ref{flowchart}). In the developed code, the total volume density of the aggregates controls the total number of aggregates generated, and the aggregate size follows a user-defined aggregate size distribution in the code. After the aggregates are generated, ITZ is created by slightly shrinking the aggregate polygon, creating a thin layer of uniform thickness around each aggregate. Finally, the code stores the coordinates of the aggregates and the corresponding ITZ points in various lists, which are then written to text files for use in subsequent steps of the pipeline.

\subsubsection{CAD file generation}
In the next step, IGES files are created which are crucial for the generation of mesh data from Abaqus/CAE. The developed Python code utilizes the geometric data generated in the previous step and implements various modules of the \texttt{OCC} python library, which provides tools for geometric and topological operations to generate the IGES CAD file. The \texttt{OCC} Python library contains tools for creating geometric primitives (points, edges, and surfaces), processing topological shapes, performing shape-splitting operations, and for exporting IGES files.

The code first constructs a square wireframe to represent the boundary of the concrete specimen. This wireframe is then used to create a face, serving as the base shape of the 2D concrete mesostructure. Next, the code generates wireframes for ellipses and the ITZ by importing the list of their corresponding constructive points generated in the previous step. A shape-splitting tool from the \texttt{OCC} library is initialized with the square frame face, preparing it for the splitting operation. The code then creates wires for each ellipse and ITZ and adds them to the shape-splitting tool. The shape-splitting tool performs the split operation on the square frame face using the added wires. The result of the split operation is utilized and added to a compound shape using the \texttt{OCC} library's tools. Finally, the compound shape is exported as an IGES file using the \texttt{OCC} library's data exchange module.
\subsubsection{Mesh file generation}

The developed Abaqus-Python code begins by importing an IGES file generated in the previous step, which contains the geometric data for the concrete mesostructure. Next, the script reads the constructive points of the aggregates and ITZ generated in the first step. The Abaqus-Python code processes this geometric data to extract the center of each aggregate particle and a point at the middle of its interface layer. These calculated points are then used to define geometrical sets representing inclusions, ITZ, and the matrix. Each set corresponds to a specific region within the concrete microstructures and is given descriptive names to facilitate later operations. Afterward, mesh generation is carried out by first defining global mesh parameters and specifying the mesh size. In this work, a mesh size with an average element length of 0.15 mm is used to discretize a 2D model with the size of $50 \, mm \times50 \, mm$, and four-node quadrilateral elements are assigned as the element type (see Fig.~\ref{2D}). An important aspect of the discretization process is to maintain continuity between the element surfaces in the different phases of the concrete microstructures. To assign the boundary conditions at the final step of the pipeline, the script defines sets for the left, right, and bottom edges. These sets will be used to apply boundary conditions in the simulation. Finally, after discretization, the code outputs an INP file. This INP file contains element and node sets for aggregates, ITZ, and the mortar matrix, as well as node sets for the main edges of the main frame.

\subsubsection{CPFM input file generation}
In the final stage of the pipeline, the generated INP file from the previous step is imported into another Python script that implements object-oriented techniques. The main goal of this part is to modify the Abaqus INP file to make it suitable for cohesive phase-field fracture simulations. The code first creates datasets of nodes, elements, and element connectivity lists and then generates three layers of elements suitable for CPFM simulations using an Abaqus UEL subroutine. These layers are suitable for the staggered solver originally developed by \citet{molnar20172d}. Furthermore, boundary conditions and material properties are applied for each phase based on the sets of elements corresponding to the different phases of the concrete and the sets of nodes representing the edges of the main frame of the specimen. Finally, the modified input file containing the Abaqus UEL subroutine is saved in a specific folder.

\subsection{Finite element simulations} \label{2nd_step}
\begin{figure}[h]
  \centering
  \begin{subfigure}{0.35\textwidth}
    \includegraphics[width=\linewidth]{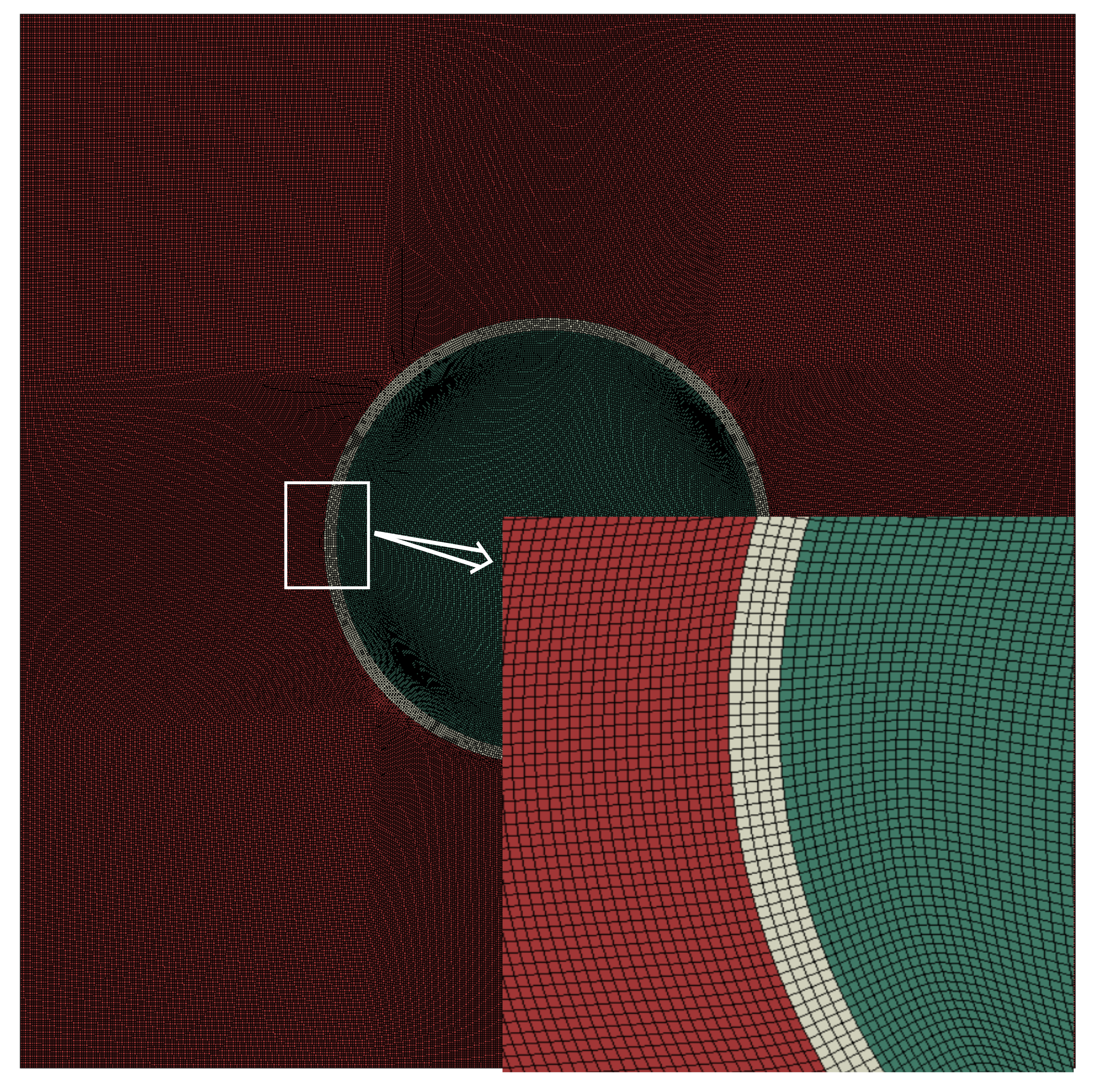}
    \caption{}
    \label{fig:figure_a}
  \end{subfigure}
  \hfill
  \begin{subfigure}{0.45\textwidth}
    \includegraphics[width=\linewidth]{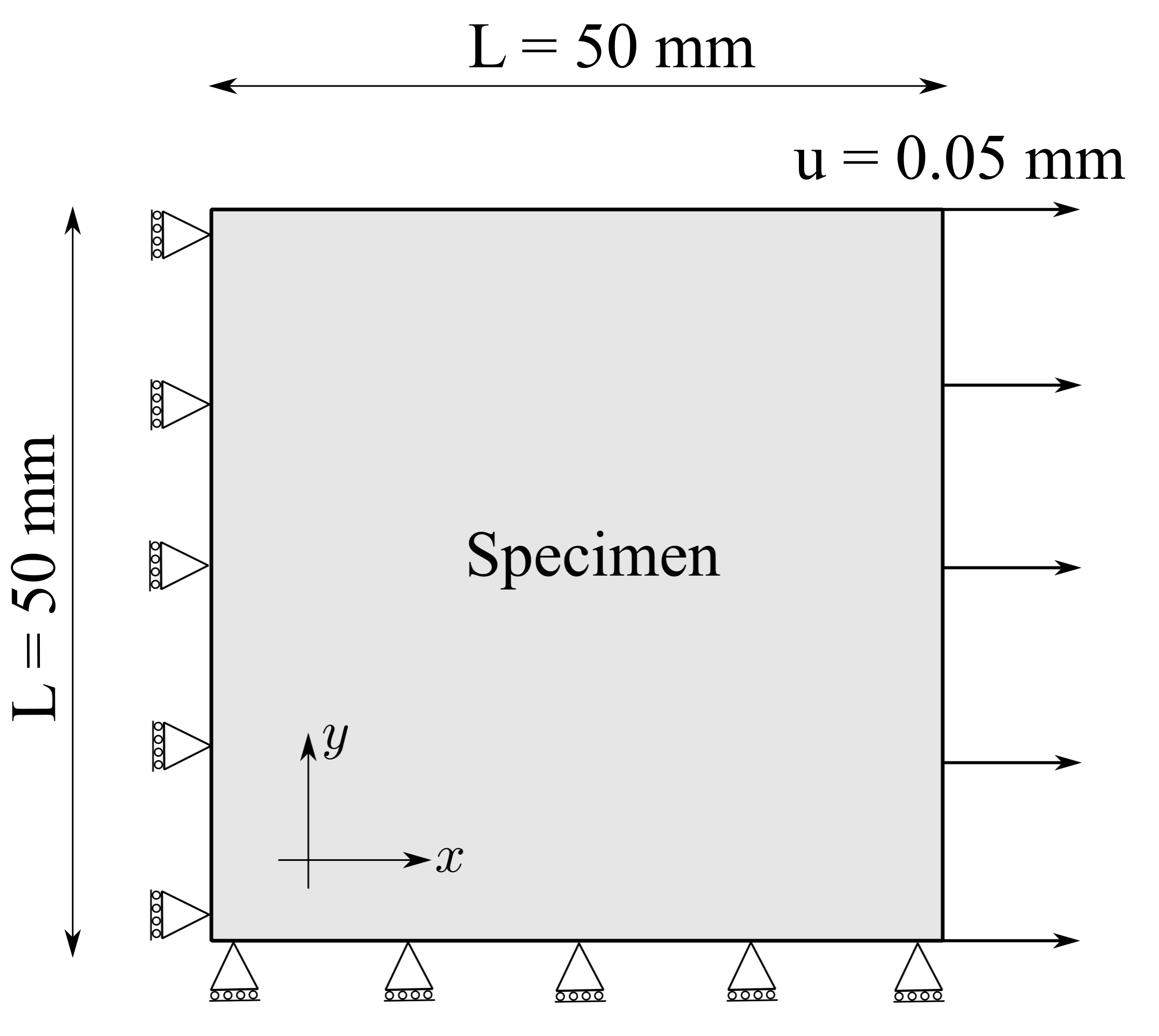}
    \caption{}
    \label{fig:figure_b}
  \end{subfigure}
  \caption{(a) represents a simple discretized concrete FE model where the white region represents the ITZ. (b) shows applied boundary conditions in all the cohesive phase-field simulations.}
  \label{2D}
\end{figure}

In the present study, the 2D model is 50 mm on each side, and the thickness of the ITZ is assigned to be 0.6 mm. A mesh size with an average element edge length of 0.15 mm is chosen for discretizing the model. This is essential to ensure that the ITZ is discretized with at least four elements over its entire thickness to allow adequate diffusion of the phase field damage within this zone. In this work, all 2D samples are subjected to a displacement of 0.05 mm in the x-direction. Figure \ref{fig:figure_b} depicts the boundary conditions applied in the finite element simulations. The staggered scheme suggested by \citet{molnar20172d} is used to solve the system of non-linear equations. Material properties relevant to the CPFM simulations are provided in Table \ref{PFM_sim}. It is worth noting that aggregate particles generally have significantly higher mechanical strength than other constituents, and thus cracks in concrete typically do not propagate into the inclusions. To account for this inherent behavior, the failure strength of inclusions is deliberately set to a much higher level (see Table.~\ref{PFM_sim}) than that of the other phases in the PFM simulations.

%To keep the effort within limits, ten FE simulations are carried out simultaneously, controlled by a Python script. The computational time for each simulation, utilizing 5 CPUs, is approximately 5 hours on a workstation equipped with Intel(R) Xeon(R) Platinum 8280 CPUs.

\begin{figure}[H]
\center
    \includegraphics[width=\textwidth]{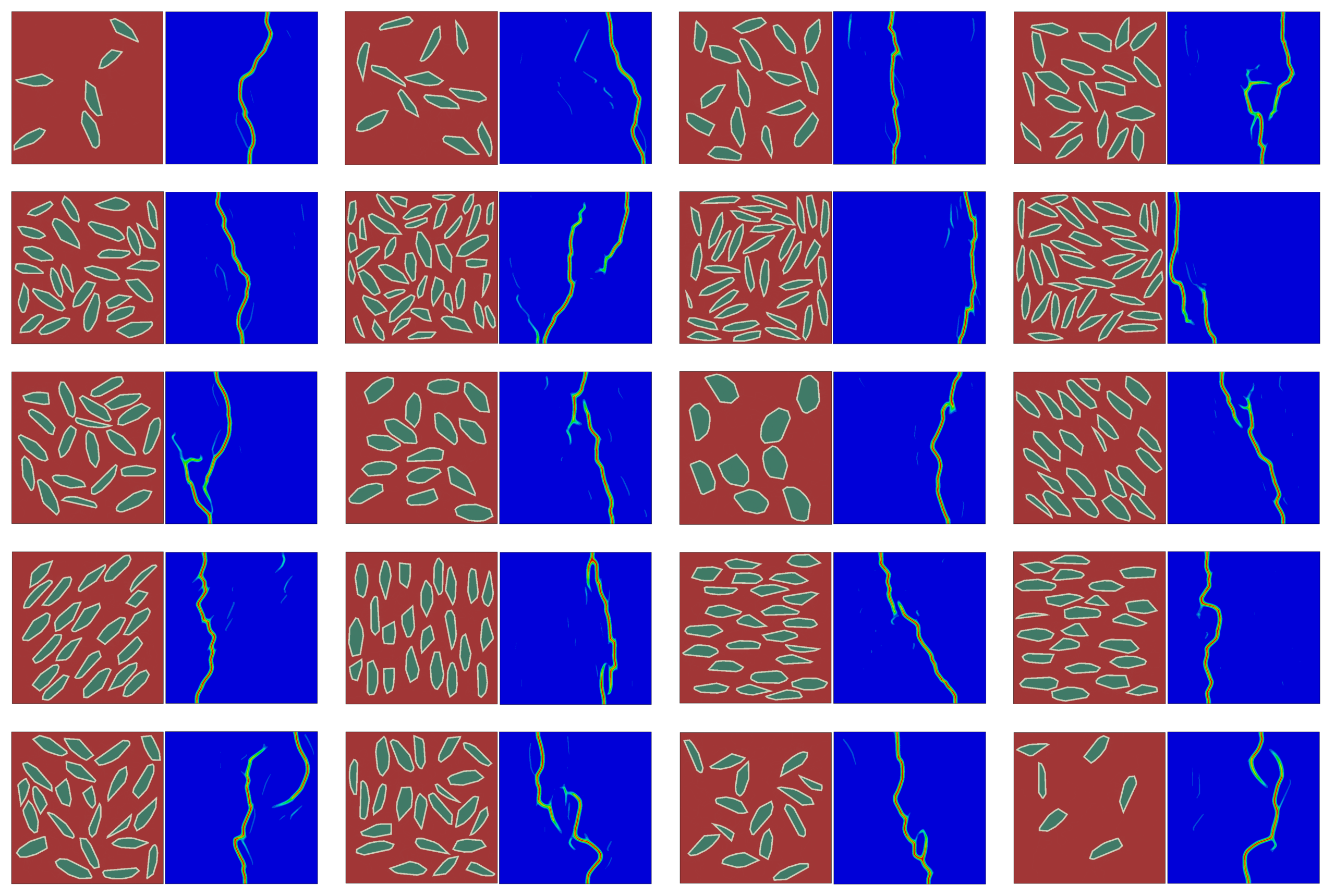}
    \caption{Cohesive phase-field fracture simulation of concrete mesostructures used to generate the required dataset for training the surrogate model. The input mesostructures contain various properties in terms of geometrical features and spatial distribution of aggregate particles.}
    \label{crack_topology2}
\end{figure}

Fig.~\ref{crack_topology2} shows some cohesive phase-field simulations of concrete mesostructure, where the simulations are carried out on models with completely varied geometrical features to ensure the diversity of the training data sets in terms of crack paths and geometrical features.
%#########################################################################
%#########################################################################
\begin{table}
\caption{Material properties assumed for phase-field simulations. The Young's modulus of the ITZ is obtained from \cite{xia2021mesoscopic}.}\label{PFM_sim}
\begin{tabular*}{\textwidth}{@{\extracolsep\fill}lcccc}
\toprule%
Phase & \makecell{Young’s modulus \\ $E$ (GPa)} & \makecell{Poisson’s ratio \\ $\nu$} & \makecell{Fracture energy \\ $G_{c}$ (N/mm)} & \makecell{Failure strength \\ $\sigma_{u}$ (MPa)}  \\
\midrule
Inclusion  & 72 & 0.16 & 0.2 & 20 \\
Matrix  & 28 & 0.2  & 0.06  & 4\\
Interface  & 21.9 & 0.2  & 0.02  & 2.4 \\
\botrule
\end{tabular*}
\end{table}

\subsection{FE results post-processing}

\begin{figure}[H]
\center
    \includegraphics[width=\textwidth]{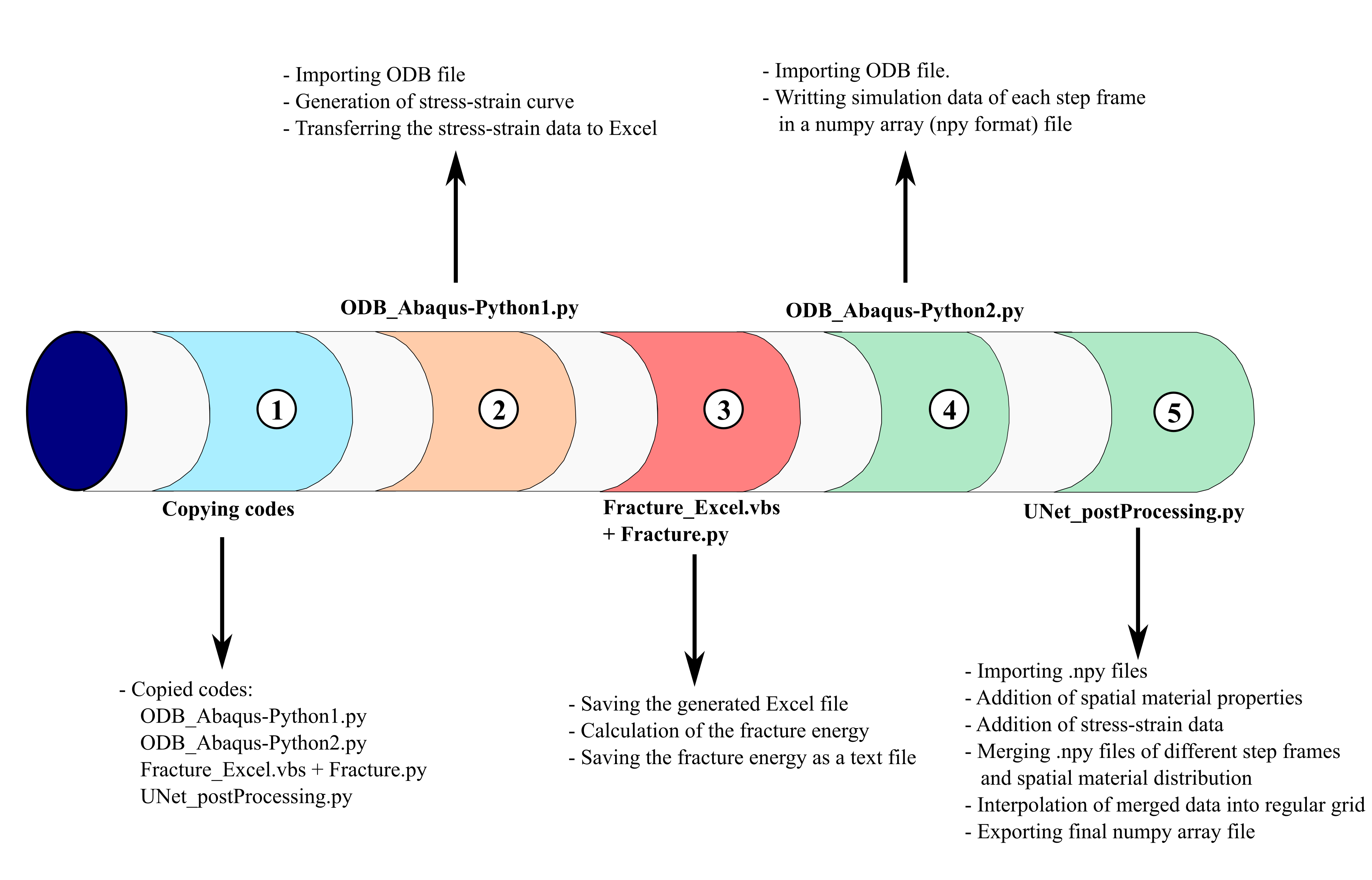}
    \caption{Fully-automated pipeline for post-processing of simulation results carried out for the mesoscale simulation of concrete materials. The outcome of the pipeline is a numpy array file for each simulation that can be used for spatiotemporal training of the neural network.}
    \label{pipeline_postprocessing}
\end{figure}

The final stage of data preparation for training the neural network involves post-processing the cohesive phase-field fracture simulations. The FE simulations performed in Abaqus/CAE are stored in a output database with file extension \texttt{.odb}. Similar to the first step, a pipeline for automatic processing of the FE results is developed in the third step.  The user only needs to specify the number of simulations to be post-processed and the processed data is saved as a NumPy array file (with the extension \texttt{.npy}) in a desired destination. As shown in Fig.~\ref{pipeline_postprocessing}, the post-processing pipeline consists of five steps, and the operations performed in each step are shown schematically. In the following sections, the operations in the post-processing phase are explained in detail.
\subsubsection{First step: Copy necessary files}
In the first stage of the pipeline, the developed codes for post-processing the FE simulation results are distributed within the simulation folders. These codes include two Abaqus-Python scripts for ODB file post-processing and two additional Python scripts for further processing the outcomes of the Abaqus-Python scripts.
\subsubsection{Second step: First ODB file automation}
In the first Abaqus-Python code, the script obtains the constituent node IDs within the left and right boundary edges of the concrete mesosample and stores them in two separate lists. Next, the stress value is retrieved along the $x-$ direction at each node in each step frame of the simulation on the left boundary nodes. The individual stress data is then summed to create a single stress data object that represents the cumulative of stress values along the $x-$ direction during the process of applying load in FE simulations. The script then generates the strain value for the strain along the x-direction at a particular node on the right edge during each step frame of the simulation. Since the displacement boundary condition is applied to the right edge, all nodes on the right edge have the same strain value. Finally, the script combines the stress-strain data and sends the data to an Excel file using the \texttt{abq-ExcelUtilities} Abaqus module.
\subsubsection{Third step: Excel file automation code}
The third step of the pipeline performs relatively simple tasks. First, it saves each generated Excel file in its corresponding simulation folder and renames it with its corresponding simulation folder name. This is done using a VBA script that interacts with Excel. The script determines the current directory path, extracts the directory name, and saves the active workbook in the current directory with the name of the directory as the file name. Next, a Python script calculates the area under the stress-strain curve, which represents the dissipated fracture energy, and saves it as a text file in the location of the simulation folder. To calculate the fracture energy, the Python code reads the saved Excel file using the Pandas library and extracts the stress and strain data from the specified columns. The function \texttt{NumPy.trapz} is used to integrate the area under the curve, which enables an accurate calculation of the fracture energy. The calculated fracture energy is then saved in a text file in the simulation folder. 
\subsubsection{Fourth step: Second ODB file automation}
Within the second Abaqus-Python code, a selective set of the FE simulation results is extracted and saved as text files at constant intervals of step frames. The complete cohesive phase-field simulation comprises 400 step frames from start to finish. In this code, the FE results, including stress, strain, displacements in the $x-$ and $y-$ direction, and the phase-field damage index for every element, are retrieved and saved every four step frames, corresponding to a constant strain value of \texttt{0.0001}. The \texttt{Centroid} technique is utilized in Abaqus-Python to efficiently extract the desired values for each element. This approach involves interpolating the results from multiple integration points to a single point at the center of the element. This technique helps achieve considerable memory savings, particularly since the cohesive phase-field fracture simulations employ small four-node quadrilateral elements with four integration points, with an average length of 0.15 mm. Considering the small element size and the overall size of the concrete mesosample, which is 50 mm, the \texttt{Centroid} approach introduces a negligible amount of error when interpolating the data for each element of the FE model to its center. 

Within the first loop of the Abaqus-Python code, the data of the first step frame is written. Since the first step frame represents the start of the simulation, the FE results for the desired parameters are all zero. However, within this first loop, the corresponding element set for each element ID is also mapped and written as a string to the output text file of the first step frame. The data from the first step frame is used in the next step of the post-processing pipeline to assign material properties to each element. In general, the modulus of elasticity, ultimate tensile strength, and fracture energy are assigned to the elements. Additionally, the coordinates of the centroid point, where the data are interpolated, are included in the first text file. Within the second loop, the script performs similar operations but writes the desired FE data along with the coordinates of the \texttt{Centroid} point for every four step frames. Therefore, with this Abaqus-Python code, the history of the cohesive phase-field simulations is stored in text files.

\subsubsection{Fifth step: Post-processing of neural network datasets }
In the final step of the post-processing pipeline, the dataset for training the neural network is extracted as numpy array files. Each numpy array file has a size of \(100 \times 8 \times 333 \times 333\). The first dimension represents the total number of step frames captured from the total of 400 step frames (one step frame out of every four is captured) from the raw FE data being processed and obtained from the previous step of the pipeline. The second dimension represents the retrieved values from the FE simulations, with the first three being material properties (modulus of elasticity, ultimate tensile strength, and fracture energy) and the remaining five being FE results (strains in the $x-$ and $y-$ directions, stresses in the $x-$ and $y-$ directions, and the phase-field damage index). The \(333 \times 333\) dimensions represent the size of the regular grid to which the FE data are interpolated.

As mentioned in Section \ref{2nd_step}, to discretize the concrete mesostructure samples, the free meshing technique is implemented in Abaqus/CAE to handle the complex mesostructure of concrete using four-node quadrilateral elements. Within the free meshing technique, considering the mesh size of 0.15 mm, the main edges of the concrete mesosample are first discretized into 333 elements, and then the discretization process propagates through the center of the concrete sample. The inner shape of discretized samples during mesh propagation is heavily dependent on the morphological properties of the concrete mesoscale samples, resulting in unique discretizations for each sample, except for the type and total number of elements on the main edges of the concrete mesosample, which is consistently 333. As a result, a possible approach for using the FE data directly for training the neural network is implementing a graph neural network (GNN) \cite{scarselli2008graph}. GNNs are well-suited for handling input data with various sizes and complex structures. In the context of concrete mesostructures, a GNN can model the relationships and interactions between different elements of the mesh more naturally than traditional neural networks. GNNs can effectively capture spatial hierarchies and dependencies by treating the mesh as a graph where nodes represent elements and edges represent their interactions. However, there are two main limitations to this approach. First, GNNs are much more complex than neural networks that use a constant size of input data. Implementing GNNs requires sophisticated data structures and algorithms to handle the varying sizes and connectivity of the input data. Second, to capture crack initiation within the ITZ, which is a very thin layer, at least four elements must be located within this area to capture the gradient terms and phase-field values accurately. Consequently, mesoscale simulations of concrete materials require a very fine mesh size. This results in raw FE data files being very large, which can create memory issues during neural network training and strain on GPU resources. 

To avoid these challenges, the fine mesh size provides the possibility to interpolate irregular FE data to regular grids with minimal error. By interpolating the data onto a regular grid, the complexity of handling variable-sized input data is reduced, making it feasible to use more straightforward neural network architectures. Therefore, in this work, to avoid the problems associated with GNNs and to account for the fine mesh size, an interpolation approach to a regular grid is chosen. This approach simplifies the data structure while preserving the necessary details for accurate analysis and modeling. In this work, we utilized \texttt{Scipy} functions to effectively transform irregular FE data into a regular grid format. Specifically, \texttt{scipy.interpolate.griddata} performs data interpolation, while \texttt{scipy.spatial.distance.cdist} and \texttt{scipy.spatial.KDTree} handles spatial distance calculations and nearest neighbor searches. The \texttt{scipy.interpolate.griddata} function maps FE simulation results—such as strains, stresses, and damage indices—onto a structured grid using methods like linear, nearest, and cubic. In this work, we implemented the cubic approach for more accuracy. The \texttt{scipy.spatial.distance.cdist} function computes distance matrices between sets of points, which is crucial for determining the closest grid points to each FE integration point and facilitating accurate interpolation.

\subsection{Analyzing errors in mapping to regular grids}
\begin{figure}[H]
\center
    \includegraphics[width=\textwidth]{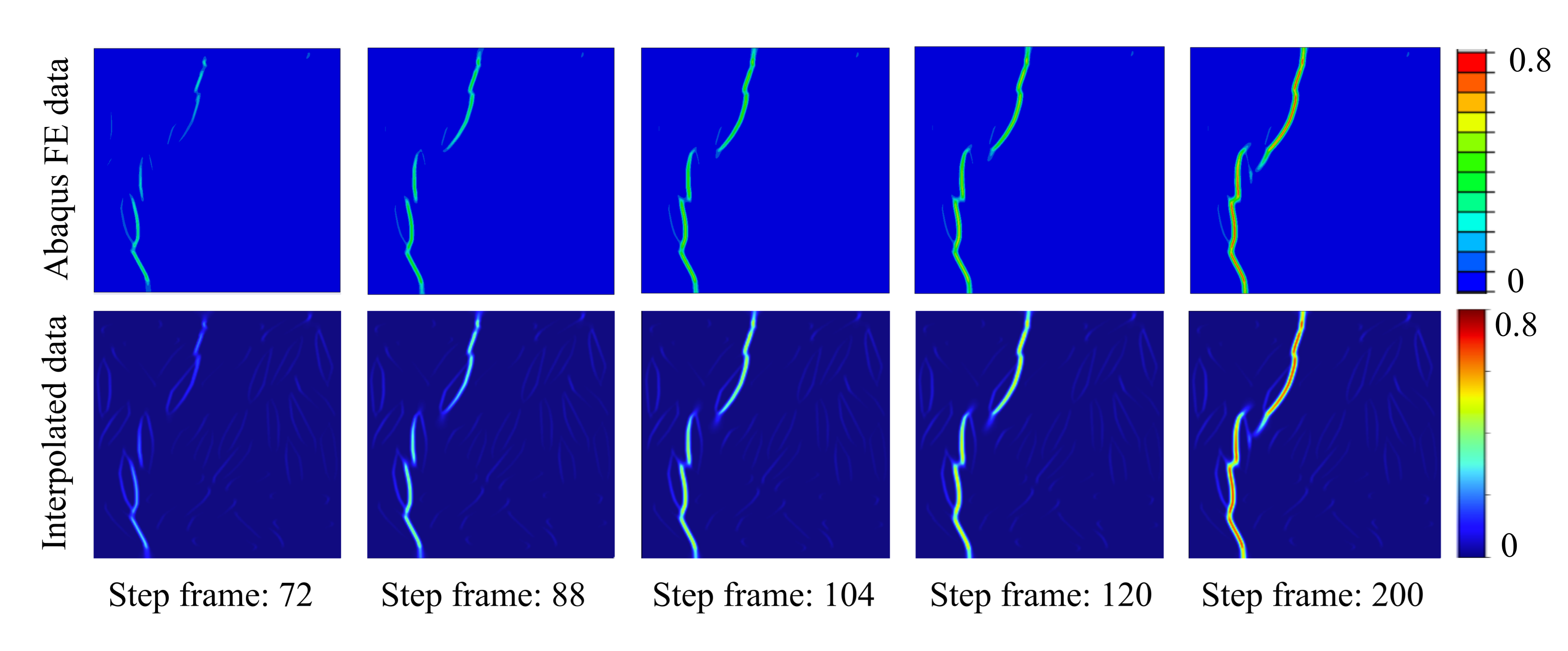}
    \caption{Interpolation of irregular Abaqus/CAE phase-field fracture simulation results onto a regular grid of size \(333 \times 333\) at various stages of damage evolution. The complete FE simulation consists of 400 step frames.}
    \label{interpolation_vs_FE}
\end{figure}

Fig.~\ref{interpolation_vs_FE} presents a comparison between the FE data results before and after the interpolation operation in terms of the distribution of the damage variable $\phi$ at different time steps. As evident, the initiation and propagation of cracks within the concrete mesoscale sample are well captured by the interpolation operation. This indicates that the unique nature of fractures in concrete microstructures, where cracks first initiate at the interface zone and then propagate in the mortar matrix, is well preserved after the interpolation operation. Therefore, the regular grids can be effectively used for training neural networks.
\section{Deep Neural Network}
In the following, the developed deep neural network for spatiotemporal prediction, as well as the training strategies, are discussed in detail.
\subsection{Developed UNet architecture}
\begin{figure}[t]
\center
    \includegraphics[width=\textwidth]{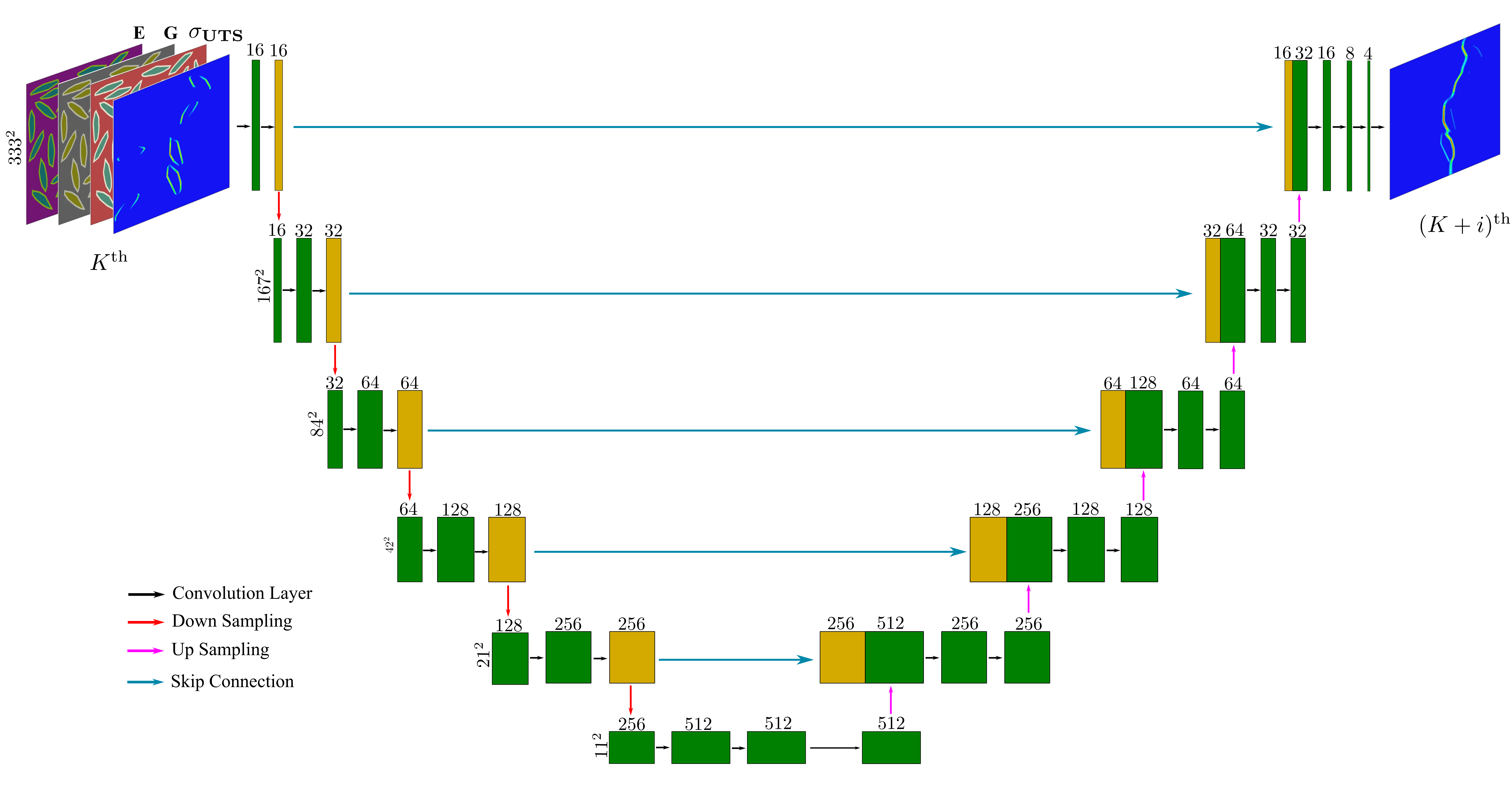}
    \caption{Implemented UNet model to predict stress and the spatial phase-field damage index in the $(k+i)$\textsuperscript{th} step frame from the input data corresponding to the $k$\textsuperscript{th} step frame. The input data has a size of \(4 \times 333 \times 333\), where the first three channels represent the spatial distribution of material properties, namely the modulus of elasticity, fracture energy, and ultimate tensile strength, and the last channel is the phase-field damage index in the $k$\textsuperscript{th} step frame.}
    \label{UNet_SS}
\end{figure}

Fig.~\ref{UNet_SS} represents the structure of the implemented UNet for the spatiotemporal prediction of crack initiation and propagation, as well as the stress-strain curve prediction in small-scale concrete microstructures. The developed UNet is characterized by its symmetric encoder-decoder structure with skip connections. Within the implemented UNet model, the input consists of the spatial distribution of the Young's modulus \(E\), fracture energy \(G\), ultimate tensile strength \(\sigma_{UTS}\), and the spatial distribution of the damage index in the $k$\textsuperscript{th} step frame. The output is the spatial distribution of the damage index in the $(k+i)$\textsuperscript{th} step frame, where the parameter \(i\) represents the jump from the input step frame to the desired prediction step frame.

In the encoder section, the model starts by processing the input image of size \(4 \times 333 \times 333\) using a \(3 \times 3\) convolutional kernel, resulting in 16 feature maps. Batch normalization and a ReLU activation function are applied to these feature maps, which retain their spatial dimensions due to padding. As the encoder progresses, each convolutional layer increases the number of feature maps while simultaneously reducing the spatial dimensions. The layers expand the feature maps from 16 to 32, followed by an increase to 128 and then to 256. In the final stages of the encoder, the number of feature maps is increased to 512, while the spatial dimensions are reduced to \(11 \times 11\).

After the final convolutional layer in the encoder section, to predict the stress value in the $(k+i)$\textsuperscript{th} step frame, the \(512 \times 11 \times 11\) feature maps are flattened, transforming the tensor into a one-dimensional array. This flattened tensor is then concatenated with the stress-strain values corresponding to the $k$\textsuperscript{th} step frame. The combined array is subsequently processed by a fully-connected feed forward neural network (FFNN). The FFNN consists of two hidden layers with 128 and 32 units, respectively, and utilizes ReLU activations between these layers. This network reduces the dimensionality from \(512 \times 11 \times 11\) to a single scalar output, which represents the stress value in the $(k+i)$\textsuperscript{th} step frame.

The decoder part aims to construct the spatial distribution of the phase-field damage index in the $(k+i)$\textsuperscript{th} step frame from the final encoded feature maps. It begins with a convolutional layer applied to the output of the encoder's final convolutional layer, maintaining the 512 feature maps and spatial dimensions. The decoder utilizes transposed convolutions for upsampling. These layers increase the spatial dimensions while maintaining or reducing the number of feature maps. At each upsampling step, the decoder concatenates the upsampled feature maps with the corresponding feature maps from the encoder, implementing the skip connections characteristic of the UNet architecture. This concatenation helps retain spatial information lost during encoding. In the first upsampling step, the spatial dimensions are doubled to \(21 \times 21\) and concatenated with the corresponding feature maps from an earlier stage of the encoder. This is followed by convolutional operations that refine the concatenated feature maps. The subsequent upsampling step further increases the spatial dimensions to \(42 \times 42\) and again concatenates the result with feature maps from a previous encoder layer. This pattern continues through multiple upsampling stages, each time increasing the spatial resolution and incorporating feature maps from the corresponding encoder layers to progressively restore the original spatial dimensions. After the final upsampling step, the decoder processes the concatenated feature maps through several convolutional operations, gradually reducing the number of feature maps. The final output, with a size of \(1 \times 333 \times 333\), represents the spatial distribution of the damage index in the $(k+i)$\textsuperscript{th} step frame.

\subsection{Developed CNN architecture}\label{CNN_S}
\begin{figure}[h]
\center
    \includegraphics[width=\textwidth]{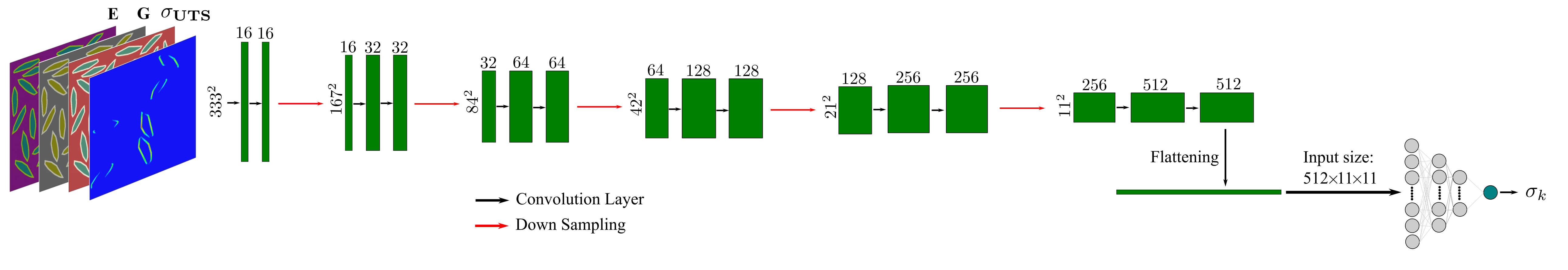}
    \caption{Implemented CNN architecture for predicting stress-strain data points at the $k$\textsuperscript{th} step frame from the input data corresponding to the same step. The input data consists of a tensor with dimensions \(4 \times 333 \times 333\), where the first three channels represent the spatial distribution of material properties alongside the phase-field damage index at the $k$\textsuperscript{th} step. This CNN architecture establishes a correlation between the spatial damage index and the corresponding stress-strain data points.}
    \label{CNN}
\end{figure}

The developed CNN model for prediction of stress-strain behavior based on spatiotemporal input data follows a structured and systematic approach for feature extraction and dimensionality reduction, as shown in Fig.~\ref{CNN}. The model starts with an input layer containing the spatial distributions of material properties such as Young's modulus (E), fracture energy (G), and ultimate tensile strength ($\sigma_{UTS}$) together with the damage index in a given step frame. The architecture goes through several convolution layers, each designed to extract and refine features from the input data. The initial convolutional layers apply \(3 \times 3\) kernels to the input data and generate feature maps, which are then processed by batch normalization and ReLU activation functions. As the model progresses, each convolution layer increases the number of feature maps and simultaneously reduces the spatial dimensions through down-sampling operations. This hierarchical feature extraction process allows the model to capture complex patterns within the input data and associate them to the average stress-strain data point. The sequence of convolutional layers expands the feature maps from an initial 4 to an eventual 512 feature maps. At the end of the coding process, the feature maps are flattened into a one-dimensional array. This flattened tensor, which contains spatial and feature information, is then fed into a FFNN. The FFNN ultimately predicts the stress value for a specific step frame. 

\subsection{UNet training phase}
To train the developed UNet model, a total of 470 datasets were utilized, with 80\% allocated for training and 20\% for validation. Due to the incorporation of three spatial distributions of material properties as input to the UNet model, the training phase required a considerably lower number of training datasets.

As mentioned earlier, in concrete microstructures, there exists a comparably small zone named ITZ with poor mechanical properties. When concrete is loaded, cracks first initiate in this zone and then propagate into other phases. Therefore, the ITZ has a considerable influence on the final crack pattern of concrete samples. To account for this critical phase in training the UNet model, a loss function specific to the ITZ is defined, focusing on calculating the loss in this region. First, a mask is generated to capture the grids corresponding to the ITZ for each training sample. The loss specific to this region is then calculated by determining the absolute error between the true and predicted values within the masked area. Additionally, at the early stages of loading, the loss value within this zone may be small. To better capture the absolute error, phase-field damage values smaller than \(10^{-10}\) are set to \(10^{-10}\). Subsequently, the phase-field damage index values are converted to the logarithmic scale using the transformation \(\phi \rightarrow 10 + \log_{10}(\phi)\). This transformation ensures that the model better captures the nuances of the damage index during training.

The batch size for both training and validation is set to 20. The \texttt{MSELoss} function is instantiated to calculate the mean squared error loss for regions other than the ITZ during training. The model is processed by a GPU for the training process and the Adam optimizer is set up with a learning rate of 0.0001 to update the model parameters. To capture a full-field prediction of the damage process, from crack initiation in the interfacial transition zone to crack bridging in the mortar matrix, a total of 10 step frame sets are captured for training, with each set undergoing 500 epochs. These step frame sets are obtained from interpolated data, ensuring that all essential stages of crack initiation, propagation, and the stress-strain curve are covered. After the training process for each step frame set, the best model's state and optimizer's state are saved and used for spatiotemporal full-field predictions. It should be noted that more step frame sets can be utilized for the training process; however, computational time and error accumulation might become significant issues. In this work, the training time for each step frame set is approximately 90 minutes using a RTX8000 GPU.

\section{Results}
The spatiotemporal surrogate models are assessed against the cohesive phase-field fracture simulation results, which are considered the \emph{ground truth}. To provide a more quantitative assessment of the accuracy of the full-field predictions at each specific time step, we use a metric based on the F1 score (Sørensen–Dice index) as suggested in \cite{mohammadzadeh2022predicting}. Finally, we compare the computational efficiency to that of the FE simulations carried out in Abaqus/CAE.
\begin{figure}[t]
\center
    \includegraphics[width=0.7\textwidth]{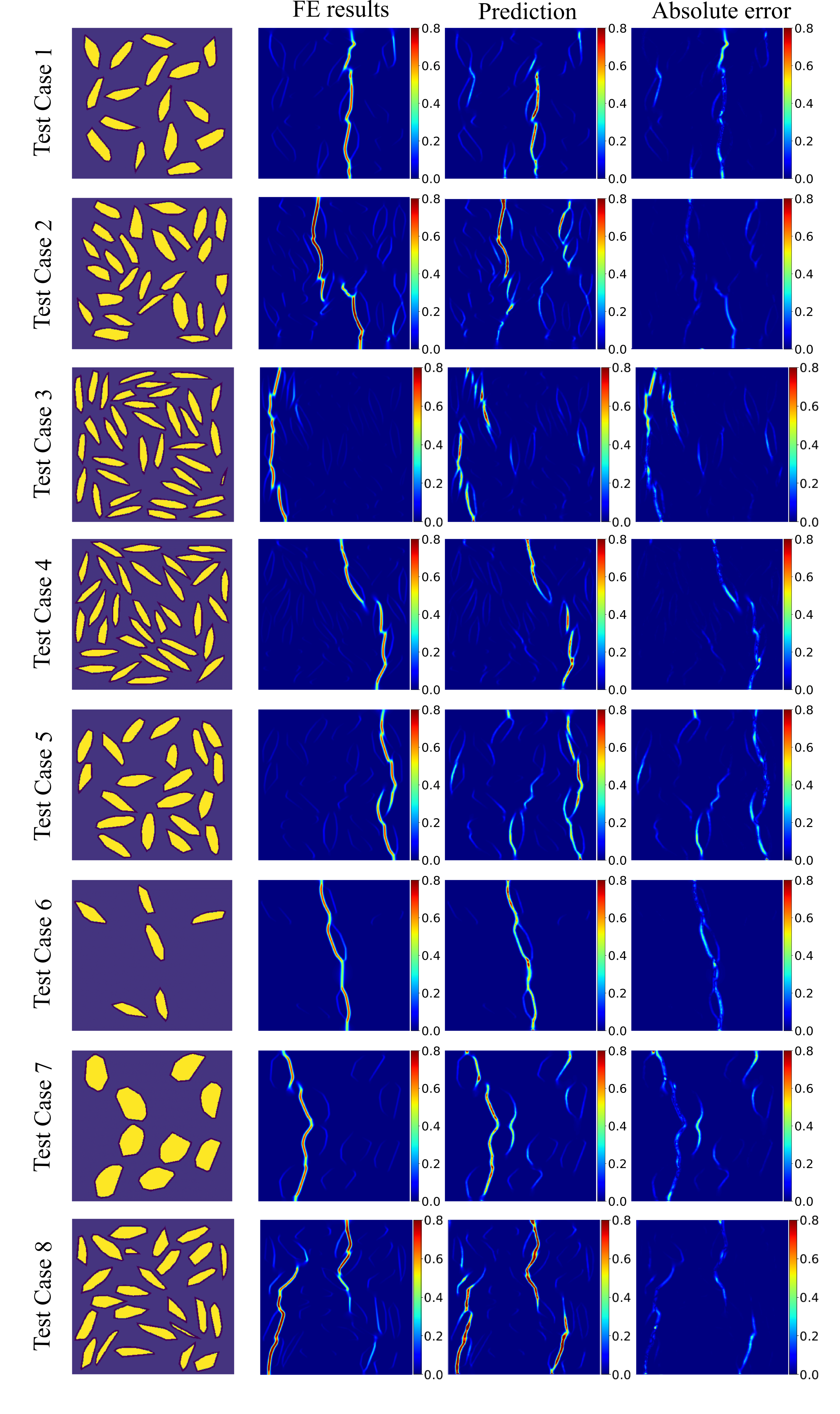}
    \caption{Predicted crack paths using the spatiotemporal full-field surrogate model on eight unseen test cases. The predictions and FE calculation results are obtained at an applied displacement of 0.015 mm. The color bar represents the phase-field damage index.}
    \label{good_predictions}
\end{figure} 
\subsection{Crack prediction results} 
Fig.~\ref{good_predictions} presents examples of predictions made by the trained spatiotemporal surrogate model on unseen test cases. These test cases involve concrete mesostructures containing three phases, which are the ITZ, mortar matrix, and the aggregate particles. The test cases exhibit various geometrical features of aggregate particles, different volume fractions of aggregates, and different angles of distribution within the mortar matrix. As observed in Fig.~\ref{good_predictions}, despite some minor deficiencies, the overall prediction of the surrogate model agrees well with the FE-calculated complex fracture paths. To facilitate a more quantitative comparison between the FE results and the spatiotemporal surrogate model predictions, we propose using the F1 score (Sørensen–Dice index). The F1 score is defined as the harmonic mean of precision and recall. Precision is the ratio of true positive results to the total predicted positives, while recall is the ratio of true positive results to all actual positives. The F1 score combines these two metrics to provide a single and robust measure of a neural network's effectiveness on unseen test cases. Mathematically, the F1 score is given by the relation

\begin{equation}
\text{F1 score} = 2 \times \frac{\text{Precision} \times \text{Recall}}{\text{Precision} + \text{Recall}},
\end{equation}
where
\begin{equation}
\text{Precision} = \frac{\text{True Positives}}{\text{True Positives} + \text{False Positives}}
\end{equation}
and
\begin{equation}
\text{Recall} = \frac{\text{True Positives}}{\text{True Positives} + \text{False Negatives}}.
\end{equation}

\vspace{2mm}

\begin{figure}[t]
\center
    \includegraphics[width=\textwidth]{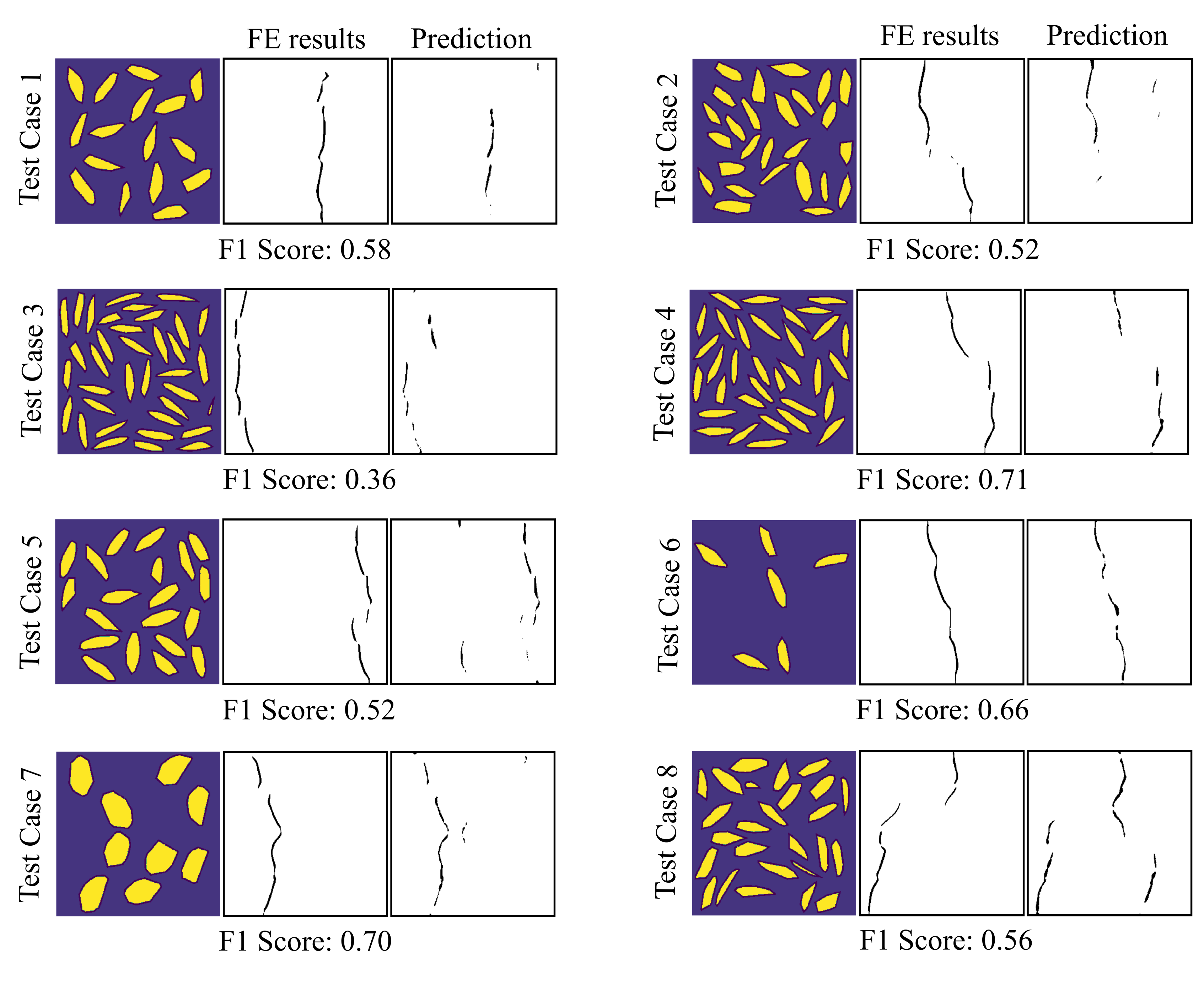}
    \caption{Quantitative evaluation of the performance of the developed spatiotemporal surrogate model in the unseen test cases using the F1 score.}
    \label{F1_score}
\end{figure}

When interpreting the F1 score, a high value (best value at 1) indicates that the model has both high precision and high recall, meaning that it is accurate in its positive predictions and actually identifies the majority of positive instances. Conversely, a low F1 score (worst score at 0) indicates deficiencies in precision, recall, or both, meaning that the model has difficulty correctly identifying positive instances or produces a significant number of false-positive predictions. To evaluate our spatiotemporal surrogate model, we calculated the F1 score to quantitatively compare the performance of the predictions of the surrogate model with the FEM simulation. The process to calculate the F1 score started with the determination of the threshold value, which corresponds to the 99th percentile of the damage index values from the FEM simulations. This threshold is used to identify regions with significant damage. Using this threshold, we segmented both the FEM simulation and the UNet prediction maps into binary images. In these segmented maps, pixels with damage index values above the threshold were marked as areas with high damage, while those below the threshold were marked as areas. 

The segmented binary maps were then flattened into one-dimensional arrays to facilitate element-wise comparison. These arrays were compared to calculate the true positives, false positives, and false negatives, which are important for calculating precision, recall, and ultimately F1 score. Using the definitions of precision and recall, we calculated the F1 score to evaluate the similarity between the segmented maps from the FEM simulations and the predictions of the developed surrogate model. As shown in Fig.~\ref{F1_score}, the spatiotemporal surrogate model on the test cases generally achieves an F1 score of approximately 0.6, demonstrating its ability to accurately reproduce crack propagation in concrete mesostructures. Additionally, Fig.~\ref{F1_score} provides examples of different F1 scores, illustrating the variations in similarity between the two maps. A detailed statistical analysis of the surrogate model's performance would be a subject of future development. 
\begin{figure}[t]
\center
    \includegraphics[width=\textwidth]{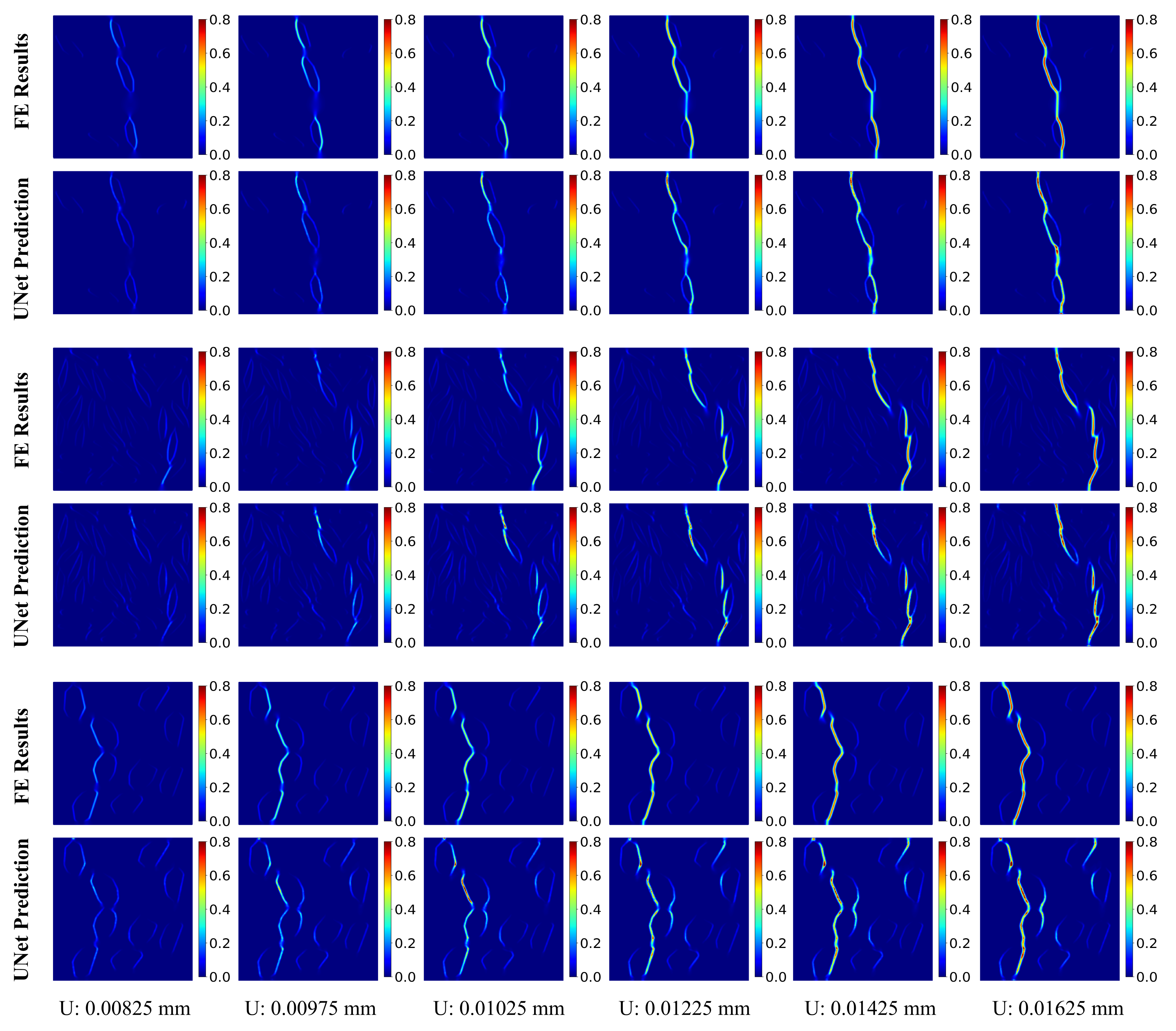}
    \caption{Evaluation of the developed spatiotemporal surrogate model against the simulated finite element data. The variable $u$ represents the applied displacement in the FE fracture simulation of the concrete mesostructure.}
    \label{history_prediction}
\end{figure}
\subsection{Stress-strain curve prediction}
The developed surrogate model demonstrates a strong capability to predict fracture propagation in concrete mesostructures. In particular, our study shows that the trained surrogate model can effectively capture the initiation of cracks in the ITZ in the early stages of loading and their subsequent propagation into the matrix. In Fig.~\ref{history_prediction}, the predictions of the model at different loading stages are compared with corresponding FE calculations. Despite some local discrepancies in the crack predictions, the overall performance of the surrogate model shows a good alignment with the FE results. In other words, the capabilities of the surrogate model include the prediction of crack initiation in the ITZ, tracking crack propagation through the matrix, and capturing complex phenomena such as crack bridging throughout the loading process.
\begin{figure}[t]
\center
    \includegraphics[width=0.8\textwidth]{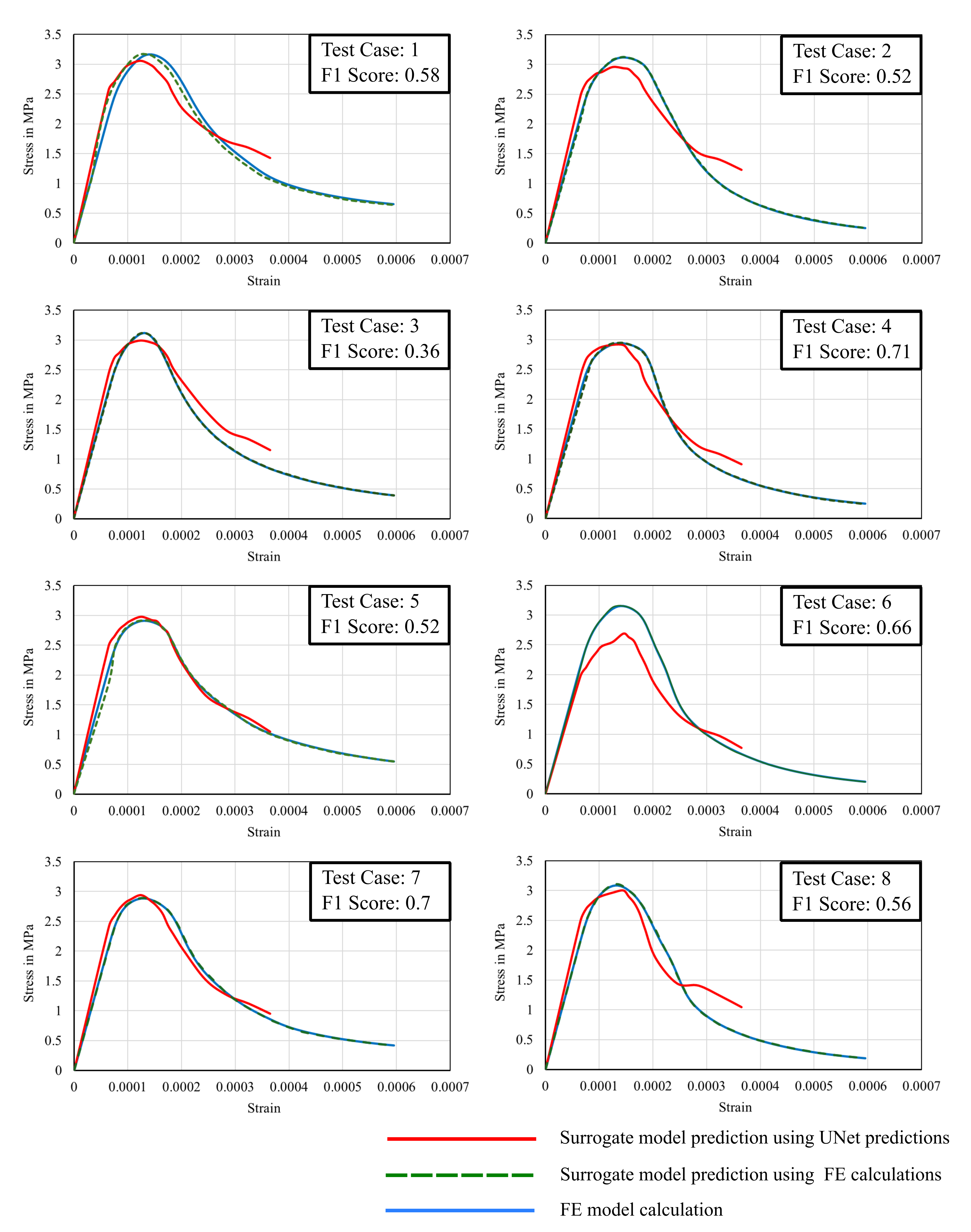}
    \caption{Predicted stress-strain curve using the trained surrogate model and its comparison with the FE calculated ones. As observed, in nearly all test cases, the maximum tensile stress, as well as the elastic and softening zones, closely match the corresponding FE calculated results. The accuracy of the predictions is higher for test cases with a higher F1 Score.}
    \label{stress_strain}
\end{figure}
Considering the capability of the surrogate model in predicting the fracture process history, a similar deep neural network based on an alternative CNN architecture was designed and trained to predict stress and strain at various stages of loading (see section \ref{CNN_S}). In this work, similar to the approached in \cite{ren2015two, naderi20223d, wang2016computational}, the stress is calculated
by dividing the total nodal reaction force of all the nodes on the reft boundary by the specimen length, while the strain is calculated as the prescribed displacement divided by the initial side
length of concrete mesostructure model, i.e., 50 mm.  In the developed CNN, similar to the surrogate model, the inputs consist of three distributions of material properties (modulus of elasticity, fracture energy, and ultimate tensile strength) alongside the spatial phase-field damage index. The output is the predicted stress-strain at that specific load step. In other words, the trained CNN establishes a correlation between the current state of the phase-field damage index in the concrete mesostructure and the corresponding point in the stress-strain relation, with the inclusion of spatial material distribution helping in an effective training process with a small number of dataset. Since the trained surrogate model provides the spatial phase-field damage index at different stages of loading, training this CNN also enables the prediction of the stress-strain curve and, consequently, the fracture energy. Fig.~\ref{stress_strain} illustrates the predicted stress-strain curve for the unseen test cases using the trained CNN compared to the FE-calculated ones. As can be observed, despite some local fluctuations, the predicted stress-strain curve provides a satisfactory accuracy in terms of the correct representation of the overall shape when compared to the corresponding FE-calculated curve. It should be noted that the accuracy of the stress-strain curve heavily depends on the surrogate model's accuracy in predicting the spatial damage index at various load steps. By increasing the F1 score of the surrogate model's predictions for the unseen test cases, the accuracy of the stress-strain curve improves correspondingly. This was tested by feeding the actual FE-calculated damage index of the test cases into the CNN instead of the surrogate model's predictions. The outcome, in this case, was a stress-strain curve that closely matched the FE-calculated one, demonstrating that the designed CNN is highly effective in correlating the spatial phase-field damage index with the stress-strain points. 

\section{Computational cost}
In Table \ref{compCost_comparison}, the computational costs associated with the test cases are presented. The FE simulation column represents the time required to perform a complete fracture simulation using a cohesive phase-field fracture approach for each test case, utilizing 2 CPUs, where each CPU has \(2.43 \times 10^8\) FLOPS (Floating-Point Operations) per core per GHz. The post-processing pipeline column represents the time taken for interpolation operations that transform the complete FE data onto a regular grid. It is important to note that the computational time for the post-processing pipeline includes the time required to interpolate the FE results for all step frames onto the regular grid. However, for utilizing the surrogate model on each test case, only the material property maps need to be transformed onto the regular grid, and the time required for this operation is approximately five minutes. Therefore, the actual post-processing time for preparing a test case as input for the surrogate model is around five minutes. The last two columns show the computation times of the surrogate and CNN models. Considering the time required, it should be mentioned that the computational cost is reduced by 2864 times, which represents a considerable reduction in computational effort.

\begin{table}[h]
\centering
\caption{Computational costs of the test cases shown in Fig.~\ref{good_predictions}. The post-processing time required to interpolate the material property maps onto a regular grid and to feed it into the surrogate model is approximately five minutes.}
\label{compCost_comparison}
\begin{tabular*}{\textwidth}{@{\extracolsep{\fill}}lccccc}
\toprule
Test cases& FE Fracture Simulation & Post-processing Pipeline & UNet Model & CNN Model\\
\midrule
Test Case 1 & 3.7 h   &  45 min  & 3.66 sec   & 1.20 sec \\

Test Case 2 &  3.4 h &  45 min   &  3.36 sec & 1.20 sec \\

Test Case 3 &  3.6 h &  45 min  &   3.78 sec & 0.57 sec \\

Test Case 4 & 3.6 h   &  45 min   & 3.84 sec & 1.22 sec\\

Test Case 5 &  3.7 h  &  45 min   &  3.54 sec  & 0.56 sec\\

Test Case 6 &  3.3 h  &  45 min   &  3.66 sec  & 0.70 sec\\

Test Case 7 &  3.7 h  &   45 min  &  3.90 sec  & 0.75 sec \\

Test Case 8 &  3.7 h   &  45 min   &  3.90 sec  & 1.10 sec\\
\midrule
\textbf{Average} & \textbf{3.58 h} & \textbf{45 min} & \textbf{0.06 min}  & \textbf{0.9 sec} \\
\bottomrule
\end{tabular*}
\end{table}

\section{Conclusion}
In this work, a surrogate spatiotemporal deep neural network based on the UNet framework is developed. This network is capable of capturing the history of crack initiation and propagation in concrete mesostructure and obtaining the average stress-strain curve. The main novelty of the developed surrogate model is that it successfully captures the initiation of cracks in the ITZ of the concrete mesostructure and their subsequent propagation in the mortar matrix. The training data is obtained by carrying out 470 cohesive phase-field simulations on concrete mesostructures, with 80\% used for training and 20\% for validation. The very low number of required data for training the surrogate model is due to the implementation of another novelty in this work. Instead of implementing image-to-image mapping, the input contains four channels where the first three represent the spatial material distribution (Young's modulus, ultimate tensile strength, and fracture energy), and the fourth is the spatial damage phase-field damage index. Due to the complexity of the concrete mesostructure, a free meshing technique is implemented for discretization purposes. Consequently, the final FE results obtained using Abaqus/CAE are stored at irregularly distributed integration points. To facilitate the implementation of simpler surrogate models, a pipeline for converting the irregular FE data to regular grids is designed. The results of the developed code and algorithms, including the pipelines and surrogate model, on the unseen test cases, and their comparison with the ground truth FE data, demonstrate the successful implementation of the novelties in this work.

Considering the possibility of converting FE data to regular grids, a potential enhancement for future development is to apply cohesive phase-field equations within the surrogate model to increase the accuracy of the predictions. Additionally, implementing neural operator learning to train such a surrogate model parametrically is another possible advancement that could eliminate constraints on specific boundary conditions \cite{liu2024multi, REN2022114399}. Finally, given that the influence of the interface zone is fully captured in the developed surrogate model and acknowledging the significant impact of this zone on fracture properties of concrete materials, microstructural optimization can also be utilized using the deep neural network. 

\backmatter

\section*{Acknowledgments}
The authors are grateful to dtec.bw – Digitalization and Technology Research Center of the Bundeswehr for their financial support. dtec.bw is funded by the European Union – NextGenerationEU.
The author Shahed Rezaei would like to thank the Deutsche Forschungsgemeinschaft
(DFG) for the funding support provided to develop the present work in the project Cluster
of Excellence “Internet of Production” (project: 390621612).

\section*{Declarations}
The authors declare no conflict of interest

\section*{Data availability}
The codes and data associated with this research are available upon request and will be published online following the official publication of the work.

\begin{appendices}

\newpage
\section{Flowchart - Mesostructure generation}\label{secA1}

\begin{figure}[H]
\center
    \includegraphics[width=1\textwidth]{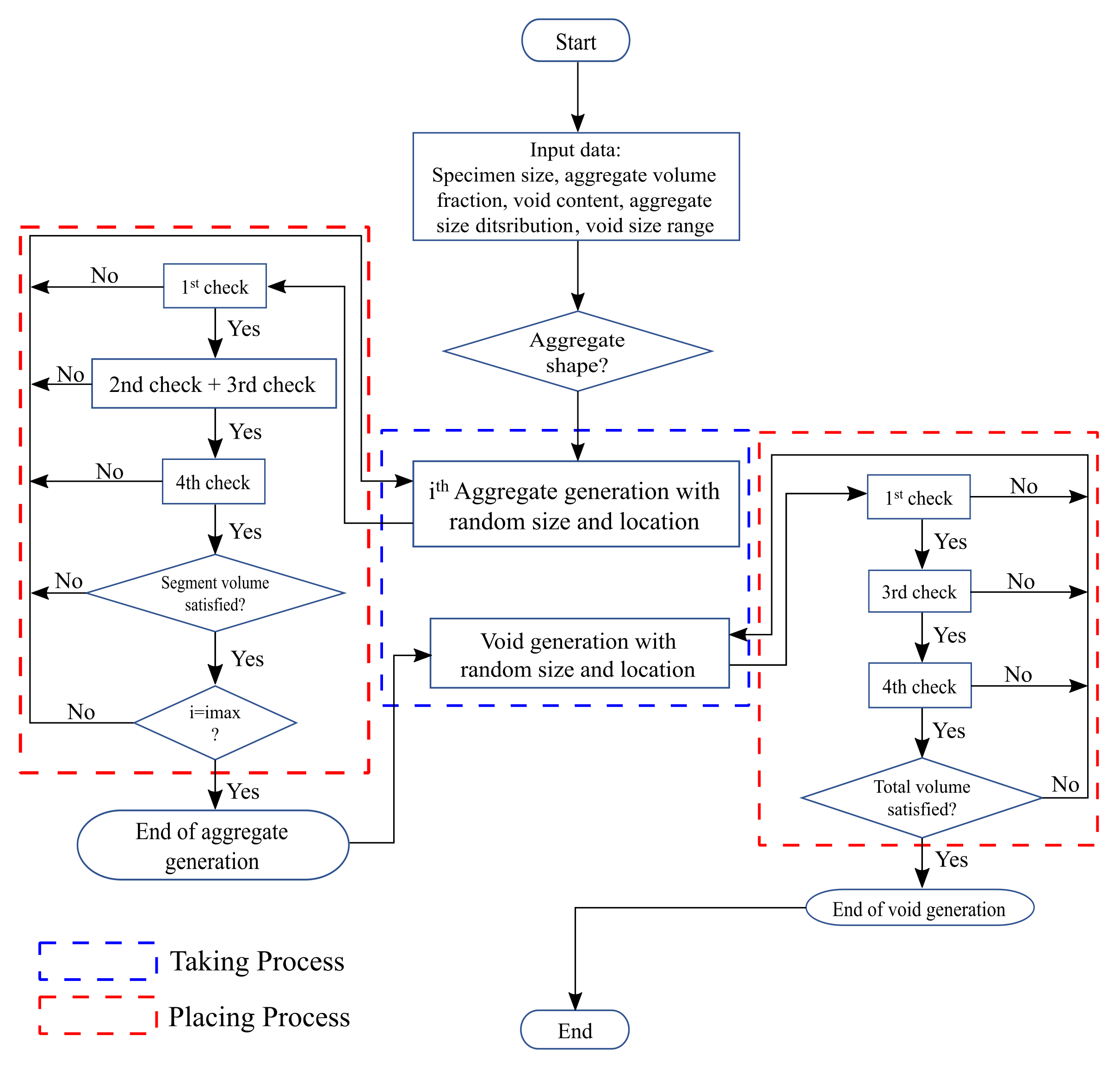}
    \caption{\label{validation}Flowchart to generate aggregates and voids \cite{najafi2023two}.}
    \label{flowchart}
  \end{figure}

\end{appendices}

\end{document}